\documentclass[%
 reprint,
 amsmath,amssymb,
 aps,
 prb,
 floatfix,
 longbibliography,
]{revtex4-2}

\usepackage{graphicx}
\usepackage[dvipsnames]{xcolor}
\usepackage{microtype}
\usepackage{braket}
\usepackage{mathtools}
\usepackage{soul}
\usepackage{comment}

\usepackage{booktabs}

\usepackage{tikz}
\usepackage{amsmath}
\usepackage{amssymb}
\usetikzlibrary{arrows.meta, decorations.markings, calc}

\usepackage[%
    colorlinks={true}, 
    linkcolor={Blue},
    citecolor={Blue},
    urlcolor={Blue},
    bookmarks={true},
    pdftitle={Fracton nonlinear response},
    pdffitwindow={true},
    pdfpagetransition={Fade},
    pdfcenterwindow={true}
]{hyperref}

\renewcommand{\vec}[1]{{\boldsymbol{#1}}}

\usepackage{mathtools}
\makeatletter
\g@addto@macro\bfseries{\boldmath}
\makeatother

\DeclarePairedDelimiter\abs{\lvert}{\rvert}%
\DeclarePairedDelimiter\norm{\lVert}{\rVert}%
\makeatletter
\let\oldabs\abs
\def\abs{\@ifstar{\oldabs}{\oldabs*}}
\let\oldnorm\norm
\def\norm{\@ifstar{\oldnorm}{\oldnorm*}}
\makeatother

\begin{document}

\title{Fingerprinting fractons with pump-probe spectroscopy}

\author{Wei-En Tseng}
\affiliation{Department of Physics and Center for Theory of Quantum Matter, University of Colorado Boulder, Boulder, Colorado 80309, USA}
\author{Oliver Hart}
\affiliation{Department of Physics and Center for Theory of Quantum Matter, University of Colorado Boulder, Boulder, Colorado 80309, USA}
\author{Rahul Nandkishore}
\affiliation{Department of Physics and Center for Theory of Quantum Matter, University of Colorado Boulder, Boulder, Colorado 80309, USA}

\date{\today}

\begin{abstract}
We demonstrate how pump-probe techniques enable specific spectroscopic diagnostics of fracton phases of matter by studying lineon-planon braiding in the paradigmatic X-cube model. Our discussion builds on previous works probing anyonic exchange statistics in conventional spin liquids, but the extension to fracton phases reveals qualitatively new phenomena due to the restricted mobility of fractionalized excitations. A key feature is that nearby planons can form an emergent bound state by accessing different planes via alternative pairing configurations. We show that this bound state qualitatively modifies the long-time linear and nonlinear responses, leading to an asymptotic linear-in-$t$ behavior for the pump-probe signal $\chi_\mathit{ZZX}$. By contrast, swapping the pump and probe polarizations produces a nonlinear response $\chi_\mathit{XXZ}$ that is asymptotically $t$-independent. This asymmetry reflects the fact that the two species of fractionalized excitations live in different spatial dimensions. Thus, the pump-probe signals studied here are sensitive to (i) nontrivial braiding statistics in three dimensions, (ii) the existence of bound states among fractionalized excitations, and (iii) the one-dimensional mobility of lineons. Our results therefore provide spectroscopic signatures that distinguish fracton phases from conventional topologically ordered spin liquids.
\end{abstract}

\maketitle


\section{Introduction}
Fracton phases of matter (see Refs.~\cite{NHreview, PretkoReview, GromovRadzihovsky} for reviews) represent an exciting frontier for condensed matter physics. They connect to numerous research programs in theoretical physics, including constrained quantum dynamics \cite{Gromov_2020, Pai_2019}, topological phases of matter \cite{Vijay_2015,VHF}, quantum error-correcting codes \cite{Haah_2011, Brown_2020}, and gravity \cite{Pretko_2017}. The connection most pertinent to the present work is the connection to topological phases. Now a key open question is: suppose a fracton phase were realized in some material -- how could we diagnose it experimentally? We are particularly interested for the present work in {\it spectroscopic} diagnostics. 

A partial answer was offered in Ref.~\cite{Nandkishore2021fingerprints}, where it was pointed out that non-linear response (accessed through the technique of two-dimensional coherent spectroscopy \cite{Mukamel}) could diagnose deconfinement (inherent in all topologically ordered phases, including fractons). However, it would be greatly desirable to have diagnostics tailored specifically for fractons. More recently, it was pointed out that nonlinear response has characteristic signatures of anyonic braiding statistics~\cite{McGinley2024signatures,Yang_PNAS,kirchner2025measuring}. Since fracton phases also have nontrivial braiding relations for the excitations~\cite{PaiHermele, Shirley_2019, Slagle_2017}, it is tempting to ask whether these techniques could be applied to the fractonic context. 

In this manuscript we demonstrate how ideas from Refs.~\cite{McGinley2024signatures,Yang_PNAS, kirchner2025measuring} can be applied to the fractonic context, by probing {\it lineon-planon braiding} in the X-cube model \cite{VHF}, a paradigmatic example of a fracton phase. We further demonstrate that the fractonic realization of these ideas differs qualitatively from previous studies due to the existence of {\it planon bound states}, which have no analog in the original discussions \cite{McGinley2024signatures,Yang_PNAS, kirchner2025measuring}. {Moreover, the nonlinear response signals exhibit a strong asymmetry under swapping the pump and probe polarizations. We summarize the results for asymptotic long-time response functions in Table\,\ref{tab:response-scaling}.
The discussion in the main text is largely intuitive and phenomenological, with technical details relegated to the Appendices. The results demonstrate that the pump-probe nonlinear response signal is sensitive to (i) the existence of nontrivial braiding statistics in a three-dimensional system, (ii) to some of the fractionalized excitations forming bound states, and (iii) to some of the fractionalized excitations being mobile in only one dimension. This combination is highly suggestive of fractionalized excitations with subdimensional mobility, a characteristic of fracton phases.

\begin{table}[t]
\centering
\renewcommand{\arraystretch}{1.18}
\resizebox{0.96\linewidth}{!}{%
\begin{tabular}{ccc}
\toprule
Response functions\; & Bound state \; & Extended states\; \\
\midrule
$\chi_Z^{(1)}(t_2)$
&
$t_2^{0}$
&
$ [t_2(\log t_2)^2]^{-1}$
\\

$\chi_\mathit{ZZX}(t_1,t_2)$
&
$t_2$
&
$ \sqrt{t_2}/(\log t_2)^2$
\\

$\chi_X^{(1)}(t_2)$
&
$-$ 
&
$t_2^{-3/2}$
\\

$\chi_\mathit{XXZ}(t_1,t_2)$
&
$-$
&
$t_2^{0}$
\\
\bottomrule
\end{tabular}}
\caption{Asymptotic long-time scaling of different response functions.
$\chi_Z^{(1)}(t_2)$ and $\chi_X^{(1)}(t_2)$ denote the linear
response functions for planon pairs and lineon pairs,
respectively. $\chi_\mathit{ZZX}(t_1,t_2)$ represents the pump-probe
response in the ZZX protocol, which captures a planon
pair braiding around a single lineon.
$\chi_\mathit{XXZ}(t_1,t_2)$ is the pump-probe response
in the XXZ protocol, which captures a lineon pair
braiding with a single planon. A pair of planons can form
a bound state that dominates the long-time behavior of $\chi_Z^{(1)}(t_2)$ and $\chi_\mathit{ZZX}(t_1,t_2)$. All results are calculated
including the hardcore constraints on the fractionalized
excitations, which significantly modify the late-time
behavior compared to the noninteracting cases.}
\label{tab:response-scaling}
\end{table}

\section{X-cube model}
Place qubits on the edges of a three-dimensional $L\times L\times L$ cubic lattice. In the X-cube Hamiltonian~\cite{CastelnovoXcube,VHF}, qubits interact according to
\begin{equation}
    H_\text{XC}(\vec{0})
    =
    - J_x \sum_{c} A_c - J_z\sum_{v, \mu} B_{v}^{\mu} 
\end{equation}
where $A_c$ is equal to the product of $X_e$ for all $e$ incident to the cube $c$, and $B_v^{\mu}$ with $\mu \in \{ x, y, z \}$ is equal to the product of $Z_e$ over the edges incident to vertex $v$ in the plane perpendicular to $\mu$, i.e., $\{ \partial e \ni v : e \perp \mu \}$. We assume that the coupling constants $J_x, J_z$ are positive.

The X-cube model has excitations with restricted mobility. Consider acting with a single $Z_e$ operator on some edge $e$. This $Z_e$ operator commutes with all $B_v^\mu$ but anticommutes with the four $A_c$ that contain the edge $e$. The four excited cubes $A_c$ are called \emph{fractons}, which are immobile when isolated. However, a pair of neighboring fractons, called a planon, is partially mobile under local operators.  Suppose that $e$ was oriented in the $x$ direction, then pairs of defective $A_c$ operators can be separated (with zero energy cost) by acting with a string of $Z_e$ operators belonging to either the $x$-$y$ or $z$-$x$ planes. Such pairs of defective $A_c$ operators are therefore mobile within a plane. We denote the excitation energy for a single planon as $\Delta^{(0)}_p = 4 J_x$.

The X-cube model also has \emph{lineon} excitations. Consider acting with a single $X_e$ operator on a $z$-oriented edge. This $X_e$ operator commutes with all $A_c$ operators but anticommutes with $B_v^\mu$ such that $v \in \partial e$ (the two vertices at the end of the edge $e$) and $\mu \perp z$, corresponding to two neighboring lineons. The lineons can be separated (with zero energy cost) along the $z$ axis by acting with $X_e$ on neighboring $z$-oriented edges. Turning a corner requires creating additional lineons so lineon excitations reside at the endpoints of rigid 1D strings.
We denote the excitation energy for a single lineon as $\Delta^{(0)}_l = 4J_z$.

We now perturb the X-cube model by applying uniform magnetic fields in the $x$ and $z$ directions with strengths $h_x$ and $h_z$, respectively. The perturbed Hamiltonian is:
\begin{equation}
    H_\text{XC}(\vec{h}) = H_\text{XC}(\vec{0}) - h_x \sum_e X_e -h_z \sum_e Z_e
    .
    \label{eqn:X-cube-perturbed}
\end{equation}
For $h_x = h_z = 0$, the excitations are eigenstates of $H_\text{XC}$ and thus remain static. Turning on nonzero $h_x$ and $h_z$ leads to hybridization of the zero-field eigenstates and lineons and planons become dynamical. 
A state containing a pair of lineons is indexed by an unordered pair of vertices
$\{ v_1, v_2 \}$. We denote the pair position by the center of mass  $R = ({v_1 + v_2})/2$ and  the relative separation $x = \abs{v_2 - v_1}$ of the two lineons. A hardcore constraint is imposed at $x=0$, since the two lineons cannot occupy the same site. For convenience, we take the system size to be infinite. The relative coordinate $x\geq1$ is then defined on a semi-infinite one-dimensional (1D) chain, while $R$ remains unconstrained. Using a Schrieffer–Wolff transformation, 
we obtain an effective 1D tight-binding Hamiltonian:
\begin{equation}
    H_{\ell}=-h_x\sum_R\sum_{x\geq1} \ket{R\pm\frac{1}{2}, x\pm{1}}\bra{R, x} 
\end{equation}
where $\ket{R\pm\frac{1}{2}, x\pm{1}}$ denotes the four states corresponding to each of the two lineons moving to the left and the right. Higher-order terms in $h_x$ generate hopping processes over larger distances. The lineon-pair eigenstates are found by Fourier transformation: $\braket{ R, x | \Psi_\ell} = \Psi_\ell(R, x) = A e^{iKR}\sin kx$ with dispersion $ E_\ell(K, k) = - 4h_x\cos(K/2)\cos k = -4h_x + \varepsilon_\ell(K, k)$.
Consequently,  the energy gap to the two-lineon continuum is modified to $2\Delta_\ell = 8J_z - 4h_x$. The density of states (DOS) resembles that of a single-particle 2D tight-binding model (on a rectangular lattice) with a van Hove singularity (VHS) at half-filling $E_\ell=0$. However, for a fixed $K$, the 1D DOS has a VHS at the band edges as $g(\varepsilon_\ell)\sim1/\sqrt{\varepsilon_\ell}$. In a finite system, $K$ and $k$ are quantized according to the boundary conditions, as discussed in Appendix \ref{Appendix_A}. 
\begin{figure}[tb]
    \centering
    \includegraphics[width=0.48\textwidth]{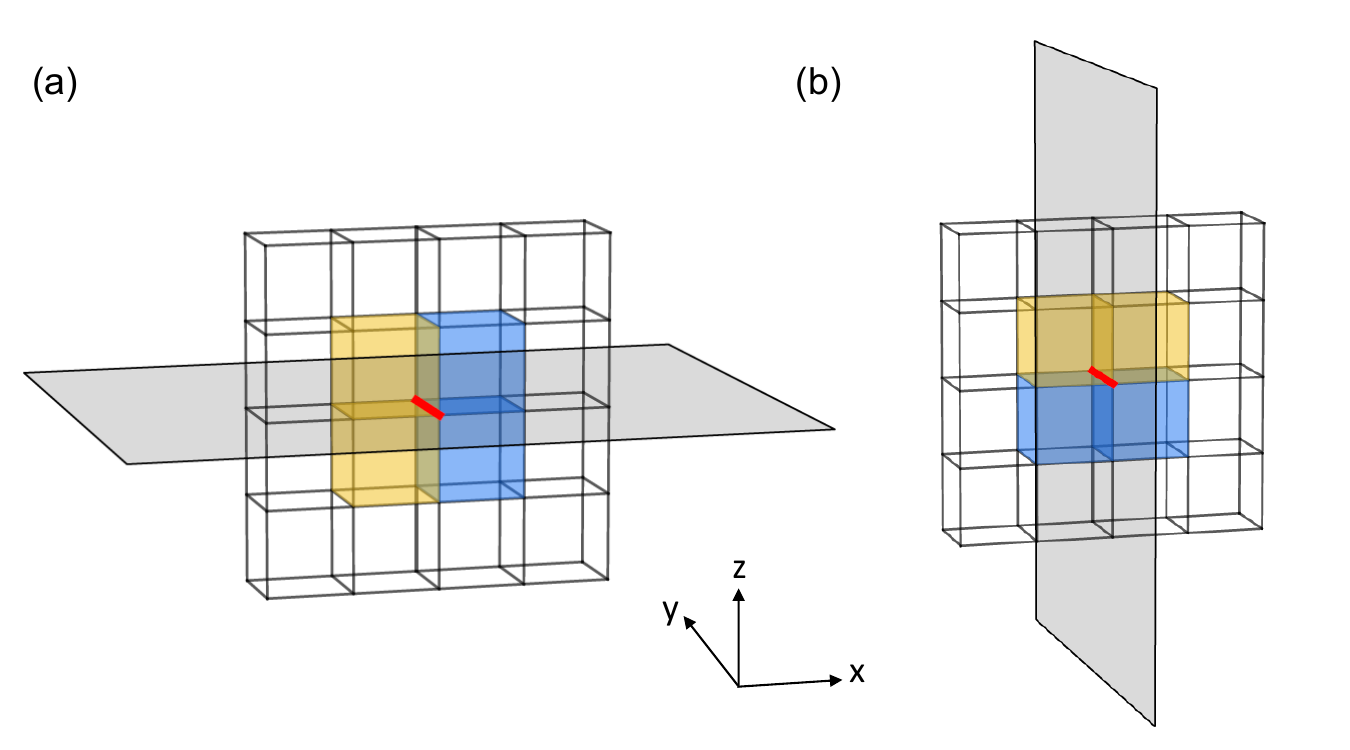}
        \caption{Different pairings of four adjacent fractons with a single $Z$ operator acting on a $y$-oriented link. (a) Planons separated in the $x$-direction can move in the $xy$-plane. This configuration is labeled by the relative coordinate as $\vec{r}_{xy}=(1,0)$. (b) Planons separated in the $z$-direction can move in the $yz$-plane. This configuration is labeled as $\vec{r}_{yz}=(0,1)$. (a) and (b) correspond to the same underlying wave function, thereby gluing together the $xy$- and $yz$-planes.}
    \label{fig:fracton_pair}
\end{figure}

\section{Planon bound state}
For a pair of planons, we write the two-particle Hilbert space in terms of a center-of-mass coordinate $\vec{R}$ and a difference coordinate $\vec{r}=\vec{r_1}-\vec{r_2}$ between the two planons.  The identical nature of the two planons implies that we identify $\vec{r}$ and $-\vec{r}$, so the planon-pair wavefunction factorizes as $\Psi_p(\vec{R}, \vec{r}) = e^{i\vec{K}\cdot \vec{R}} \phi(\vec{r})$, where $\phi(\vec{r}) = \phi(-\vec{r})$. 
Importantly, when four fractons sit on adjacent cube centers, they can be paired up into planons in different ways, as shown in Fig.~\ref{fig:fracton_pair}. Consider the action of a single $Z_e$ operator oriented along $\hat{y}$. This configuration can be viewed as a planon pair separated in the $x$-direction ($\Delta x$=1),  which is allowed to move in the $xy$ plane under the perturbed Hamiltonian  $H_{\text{XC}}(\vec{h})$. We label this configuration by the difference coordinate $\vec{r}_{xy}=(1,0)$ in the $xy$-plane. Alternatively, the same four-fracton configuration can be regarded as a planon pair separated in the $z$ direction ($\Delta z$=1), which is allowed to move in the $yz$ plane, and we label it as $\vec{r}_{yz}=(0,1)$. These two points describe exactly the same configuration, which glues the $xy$ and $yz$ planes together. The fact that planons can move in either the $xy$ or $yz$ planes, depending on how they are paired, means that they are not actually confined to move in a plane unless they are well separated. Nevertheless, one can still obtain an effective two-dimensional model by projecting out extra degrees of freedom. Focusing on the $xy$-plane by integrating out the $yz$ plane, we expect to obtain an effective local potential at $\vec{r}_{xy}=(1,0)$. Since we identify $\vec{r}$ and $ -\vec{r}$, the same local potential should also exist at $\vec{r}_{xy}=(-1,0)$. Similarly, the action of a single $Z_e$ operator oriented along $\hat{x}$ creates a planon pair labeled by $\vec{r}_{xy}=(0,1)$ and $\vec{r}_{xz}=(0,1)$, which glues the $xy$- and $xz$-plane. Projecting out the $xz$-plane results in local potentials at $\vec{r}_{xy}=(0,\pm1)$.  In addition, the two planons cannot reside on the same site, leading to a hardcore constraint at $\vec{r}_{xy}=(0,0)$, modeled as an infinite on-site potential at the origin. Finally, the effective 2D tight-binding model for a planon pair can be expressed as: 
\begin{align}
H_{p} = &-h_z \sum_{\vec{r}} \sum_{\vec\delta = \pm \hat{x},\, \pm\hat{y}}
 \lvert \vec R+\vec\delta/2, \vec{r} + \vec\delta \rangle \langle \vec R, \vec{r} \rvert \nonumber
\\ &+ \sum_{\vec{r}} V(\vec{r}) \lvert \vec R , \vec{r} \rangle \langle \vec R ,\vec{r} \rvert ,
\label{eqn:planon_hamiltonian}
\end{align}
with
\begin{equation}
V(\vec{r}) =
\begin{cases}
\lambda\rightarrow\infty, & \vec{r} = (0,0), \\
U(E), & \vec{r} = (\pm1,0);(0,\pm1) \\
0, & \text{otherwise}.
\end{cases}
\end{equation}
where we have written the relative coordinate as $\vec{r} = \vec{r}_{xy}$. In the absence of $V(\vec r)$, the two-planon eigenenergies (expressed in terms of COM momentum $\vec K$ and relative momentum $\vec k $) are $E(\vec K, \vec k) = -4h_z[\cos (K_x/2)\cos(k_x)+\cos(K_y/2)\cos(k_y)]$ which spans $-8h_z \leq E \leq 8h_z$. In the spatially uniformly excited case,  we set $\vec K=0$, and the bulk DOS has the same form as that of a 2D square lattice.  Since $V(\vec r)$ is nonzero on only a finite set of sites, it does not modify the bulk continuum DOS in the thermodynamic limit, although it does modify the scattering eigenstates. In addition, the energy-dependent local potentials are attractive $U(E)<0$ below the two-planon continuum, which generate an additional planon-pair bound state outside the continuum. Detailed analysis of the planon's bound state and extended states are presented in Appendix \ref{Appendix_B} and \ref{Appendix_D}, respectively.


\section{Pump-probe response}
\subsection{ZZX protocol}
The coexistence of bound and extended planon states can provide a distinctive fingerprint of fracton order in pump-probe spectroscopy. We first consider a protocol in which an $x$-polarized pump pulse applied at $t=0$, coupling to $\sum_e X_e$ and creating a pair of lineons. Subsequently, a $z$-polarized probe is applied at delay time $t_1$, coupling to $\sum_e Z_e$ thereby creating a pair of planons. Finally, we measure the resulting planons' response at time $t_1+t_2$. We denote this as the ZZX protocol, where the last index denotes the pump polarization and the first two indices denote the measured response and the probe polarization, respectively. The nonlinear pump-probe signal $\chi_\mathit{ZZX}$ is then extracted by subtracting the linear (unpumped) response,
\begin{equation}
    L^3 \chi_\mathit{ZZX}(t_1,t_2)
    =
    \langle {A(t_1{+}t_2) \, A(t_1)} \rangle_{\ell}
    -
    \langle {A(t_1{+}t_2) \, A(t_1)} \rangle_{0}
    ,
    \label{eq:chi-def}
\end{equation}
where $A = \sum_e Z_e$ and $L^3$ is a normalization volume.  Here $\langle {\cdots} \rangle_{0}$ and  $\langle {\cdots} \rangle_\ell$ denote expectation values in the ground state and in the two-lineon state created by the pump, respectively.  
We consider that the only interaction between the planons and the lineons is statistical with a phase factor $(-1)^{n_\ell}$, where $n_\ell$ is the number of lineons (of the appropriate flavor) enclosed by the loop taken by the planons \cite{PaiHermele}. The subtraction in Eq.~\ref{eq:chi-def} removes processes already present in linear response, so that only processes involving nontrivial braiding statistics between the pumped lineons and the probed planons contribute to the nonlinear response. Microscopically, the statistical phase arises when the string operator of the lineon pair intersects the loop operator of the planon pair. In the 3D X-cube model, the braiding processes occur when both the planons and lineons are moving in the same plane, reducing it to a 2D problem. This can be seen by considering a lineon pair moving along $\hat{x}$, i.e. a string operator of $X_e$ acts on the $x$-oriented links. If the planon pair moves in the plane perpendicular to the lineon axis  (the $yz$-plane), then the loop operators of $Z_e$ only act on the $y$- and $z$-oriented links, and no nontrivial phase is generated because there are no intersections on the same link. As a result, braiding in the X-cube model can be thought of as decoupled 2D layers.

To quantify the braiding statistics, we follow the approach of Ref.~\cite{McGinley2024signatures}. The pumped lineons move in one dimension along semiclassical trajectories $ \vec x(t)= \pm \vec vt + \vec x_i$ with characteristic velocity $\vec v = v \hat x $ and initial position $\vec x_i$. In a spacetime picture, the pumped lineon traces out a straight world-line, and the planon pair with two trajectories $\vec r_1(t)$ and $\vec r_2(t)$ forms a closed loop between $t_1$ and $t_1{+}t_2$. A topological phase is acquired only when the lineon world-line intersects the planon's world-loop, as shown in Fig.~\ref{fig:Braiding}. We denote the braiding probability as $P[\vec r_{1,2}(t);v]$ for a given pair of planon trajectories $[\vec r_1(t),\vec r_2(t)]$ and lineon velocity  $v$. The pump probe signal is then proportional to the planon's linear response $\chi_Z^{(1)}$ and the braiding probability with additional weighting: $\chi_\mathit{ZZX}(t_1,t_2)\propto \chi_Z^{(1)}(t_2) [(-1)^{n_l}-1]P[\vec r_{1,2}(t);v]$. At long delay times $t_1$, the two lineons created by the pump are typically far separated, so the subsequent probe planon pair braids only one of them, and we set $n_l=1$.

\begin{figure}[t]
\centering
\resizebox{\columnwidth}{!}{%
\begin{tikzpicture}[
    x={(0.7cm,0cm)}, 
    y={(0.45cm,0.25cm)}, 
    z={(0cm,0.85cm)},
    >=Stealth,
    line cap=round, line join=round,
    every node/.style={scale=0.99}
]

\pgfmathsetmacro{\tOne}{1}
\pgfmathsetmacro{\tTwo}{3}
\pgfmathsetmacro{\tFinal}{\tOne + \tTwo}
\pgfmathsetmacro{\vShift}{2} 

\coordinate (Ri) at (0, 3, \tOne);
\coordinate (Rf) at (0.2, 3, \tFinal);

\coordinate (Xi) at (4, 3, 0);

\coordinate (x1) at (0, 2.5, \tOne + 0.75);
\coordinate (x2) at (0, 2, \tOne + 1.6) ;
\coordinate (x3) at (0, 2.4, \tOne + 2.25) ;
\coordinate (x4) at (0.2, 3.5, \tOne + 0.75);
\coordinate (x5) at (0.2, 4, \tOne + 1.2) ;
\coordinate (x6) at (0.2, 3.5, \tOne + 2.25) ;

\pgfmathsetmacro{\y}{1}
\pgfmathsetmacro{\v}{2}
\coordinate (pRi) at (2+\vShift, 1.0+\y, 0);
\coordinate (pRf) at (2+\vShift, 3+\y, 0);

\coordinate (px1) at (1-\v+\vShift, 1.6+\y, 0);
\coordinate (px2) at (1-\v+\vShift, 2.3+\y, 0) ;
\coordinate (px3) at (3+\v+\vShift, 1.6+\y, 0);
\coordinate (px4) at (3+\v+\vShift, 2.3+\y, 0);

\draw[red!70!black,  line width=1.5pt,
      decoration={markings,
        mark=at position 0.35 with {\arrow{>}},
        mark=at position 0.80 with {\arrow{>}}},
      postaction={decorate}]
  plot[smooth] coordinates {(Ri) (x4) (x5) (x6) (Rf)};

V= 3
\path[fill=red!15, draw=red!70!black, dashed, thick] 
    (pRi) 
    .. controls (px1) and ( px2) .. ( pRf)
    .. controls (px4) and ( px3) .. ( pRi) 
    -- cycle;
    
\draw[|<->|, thin] (2.5,1.5, 0) -- (6.5,1.5, 0) 
    node[midway, below, font=\scriptsize] {$|\Delta x_{\parallel}| $};

\draw[|<->|, thin] (6.5, 1.6, 0) -- (6.5, 3.4, 0) 
    node[midway, sloped, below, font=\scriptsize, yshift=-2pt] {$|\Delta x_{\perp}|$};

\draw[->, thick] (-0.2,0,0) -- (9.5,0,0) node[below] {$x$};
\draw[->, thick] (0,-0.2,0) -- (0,4.5,0) node[right] {$y$};
\draw[->, thick] (0,0,-0.2) -- (0,0,5.5) node[left] {$t$};

\draw[thick] (0,0,\tOne) -- (-0.15,0,\tOne) node[left] {$t_1$};
\draw[thick] (0,0,\tFinal) -- (-0.15,0,\tFinal) node[left] {$t_1 + t_2$};

\draw[thin] (0,0.15,\tOne) -- (0.15,0.15,\tOne) -- (0.15,0,\tOne);
\draw[thin] (0,0.15,\tFinal) -- (0.15,0.15,\tFinal) -- (0.15,0,\tFinal);

\draw[dashed, thin] (0,0,\tOne) -- (Ri);
\draw[dashed, thin] (0,0,\tFinal) -- (Rf);

\def\aL{3.95}\def\vL{-1.5}
\def\aR{2}\def\vR{1.5}

\draw[blue!75!black, line width=1.5pt,
      decoration={markings, mark=at position 0.73 with {\arrow{>}}},
      postaction={decorate}]
  (Xi) -- ($(Xi) + ({\aL*\vL},0,{\aL})$);

\draw[blue!75!black, line width=1.5pt,
      decoration={markings, mark=at position 0.60 with {\arrow{>}}},
      postaction={decorate}]
  (Xi) -- ($(Xi) + ({\aR*\vR},0,{\aR})$);

\def\alphaLine{0.5}
\coordinate (px1) at (1.78, 2.8, 0);
\draw[black!75!black, thick, dashed, draw opacity=\alphaLine,
      postaction={decorate}]
  (px1) -- ($(px1) + 1.12*({\vL},0,{1})$);

\coordinate (px1) at (6.2, 3, 0);
\draw[black!75!black,thick, dashed, draw opacity=\alphaLine, 
      postaction={decorate}]
  (px1) -- ($(px1) + 3.95*({\vL},0,{1})$);

\coordinate (px1) at (3.5, 4, 0);
\draw[black!75!black, thick, dashed, draw opacity=\alphaLine, 
      postaction={decorate}]
  (px1) -- ($(px1) + 2.22*({\vL},0,{1})$);
  
\coordinate (px1) at (4, 2, 0);
\draw[black!75!black, thick, dashed, draw opacity=\alphaLine, 
      postaction={decorate}]
  (px1) -- ($(px1) +2.656*({\vL},0,{1})$);

\fill[blue!85!black] (Xi) circle (2pt);

\node[blue!85!black, right, font=\small] at (4.2, 2.8, 0) {$\vec{x}_i$};

\node[blue!75!black, font=\small] at (4.5, 2.0, 1.2) {$\vec{x}_i \pm \vec{v}t$};

\node[red!75!black, font=\small] at (0, 1.1 , 2.8) {$\vec{r_1}(t)$};

\node[red!75!black, font=\small] at (0.7, 4 , 2.45) {$\vec{r_2}(t)$};

\draw[red!70!black,  line width=1.5pt,
      decoration={markings,
        mark=at position 0.35 with {\arrow{>}},
        mark=at position 0.80 with {\arrow{>}}},
      postaction={decorate}]
  plot[smooth] coordinates {(Ri) (x1) (x2) (x3) (Rf)};

\draw[blue!75!black, line width=1.2pt,
      decoration={markings, mark=at position 0.80 with {\arrow{>}}},
      postaction={decorate}]
      (6,0,5.2) -- (7.5, 0,5.2)
      node[right , font=\normalsize] {Lineons};
\draw[red!75!black, line width=1.2pt,
      decoration={markings, mark=at position 0.80 with {\arrow{>}}},
      postaction={decorate}]
      (6,0,4.6) -- (7.5, 0, 4.6)
      node[right , font=\normalsize] {Planons};
  
\end{tikzpicture}
}
    \caption{Spacetime diagram of a ZZX pump-probe protocol for braiding statistics in the X-cube model. A pair of lineons is created at $t=0$, which move in one dimension along the trajectories $\vec x=\pm \vec vt + \vec x_i$, where $\vec v =v\hat{x}$. A pair of planons is created and annihilated at times $t_1$ and $t_1+t_2$, respectively, with trajectories labeled by $\vec r_1(t)$ and $\vec r_2(t)$ that together form a loop. The red shaded region indicates the two-dimensional area spanned by the initial lineon positions $\vec x_i$ for which one of the lineon world-lines intersects the planon world-loop. The braiding probability is therefore proportional to this area, which can be estimated as $A[\vec r_{1,2}(t); v] \sim \Delta x_{\parallel} \Delta x_{\perp}$, where $\Delta x_{\parallel}$ and $\Delta x_{\perp}$ denote the widths parallel and perpendicular to the lineon velocity.}
    \label{fig:Braiding}
\end{figure}
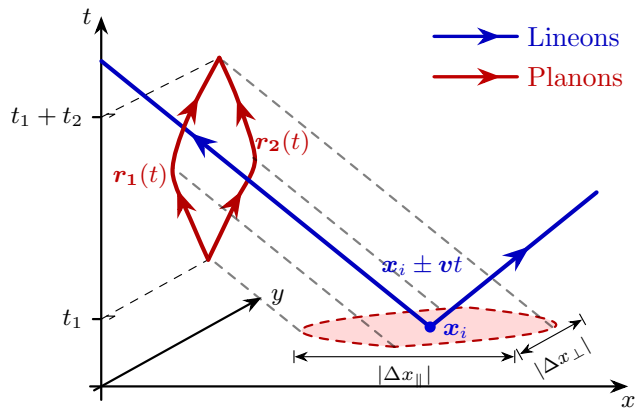


With the assumption of a homogeneous pump pulse, the initial position of the pump lineon $ \vec x_i$ is uniformly distributed in space. The braiding probability $P[\vec{r}_{1,2}(t);v]$ can then be expressed in terms of an effective area  $A[\vec r_{1,2}(t); v]$ spanned by the initial lineon positions $\vec x_i$, within which the lineon world-line intersects the planon world-loop. The time-dependence of the area can be estimated by separating the parallel width $\Delta x_\parallel$ and the perpendicular width $\Delta x_\perp$ to the lineon velocity $\vec{v}$. The perpendicular width is controlled by the typical separation of the two planons in the direction perpendicular to the lineon, $|\vec{r}_{1,\perp}(t)-\vec{r}_{2,\perp}(t)|$. Due to the quantum broadening of the planon's extended states, the perpendicular width scales as $\Delta x_\perp\propto \sqrt{h_zt_2}$ in both non-interacting and interacting cases, as shown in the Appendix \ref{Appendix_E}.
In contrast, a bound pair of planons does not spread in space, so $\Delta x_\perp\sim O(1)$ remains a constant. On the other hand, the parallel width is classical: a lineon sweeps a length $|v|t_2$ along its axis between $t_1$ and $t_1+t_2$, and shifting $x_\parallel$ is equivalent to shifting the intersection time between the lineon and the planon loop. Thus, $\Delta x_\parallel$ can be approximated as $|v| t_2$. We note the spreading of the extended planon states, which scales as $\sqrt{h_zt_2}$, is subleading compared to $\Delta x_\parallel$. Therefore, $\Delta  x_\parallel\sim|v| t_2 +O(\sqrt{h_zt_2})$ for the planon extended states.
To leading order, the area of initial conditions that yields braiding scales as  $A[\vec r_{1,2}(t); v]\approx \Delta x_\parallel \Delta x_\perp\propto |{v}|t_2 \times \sqrt{t_2}$ for the extended states and $A[\vec r_{1,2}(t); v]\propto |{v}|t_2$ for the bound state. By combining the braiding probabilities with the linear response signal, the pump-probe signal scales as

\begin{align}
\chi_\mathit{ZZX}(t_1,t_2)&\propto (-2)\chi_Z^{(1)}(t_2) \times  P\big[\bm r_{1,2}(t);v] \\
&\propto
\begin{cases}
    \chi_{e}^{(1)}(t_2) |v| t_2^{3/2} , &\text{extended states}\\[2pt]
   \chi_b^{(1)}(t_2) |v| t_2,   &\text{bound state}.
\end{cases} 
\label{eqn:pump_probe_scaling}
\end{align}
where we have decomposed the planon's linear-response susceptibility  $\chi_Z^{(1)}(t_2)$ into contributions from extended states, $\chi_e^{(1)}(t_2)$, and from the bound state, $\chi_b^{(1)}(t_2)$, since they have distinct time scaling. 

The planon's linear response can be calculated by the correlation function with respect to the ground state:
\begin{align}
    \langle A(t_1 + t_2) A(t_1) \rangle_0  &= L^3 e^{-i2\Delta^{(0)}_pt_2} \braket{2_p | e^{-iH_pt_2} | 2_p}
    \label{eqn:correlation}
\end{align}
where $\ket{2_p}$ is the normalized, translation-invariant
superposition of neighboring planon pairs, and $2\Delta_p^{(0)}$
is the unperturbed energy cost to create the two-planon state. For translation-invariant perturbations, the evaluation of the correlation function reduces to a single-particle problem in the relative coordinate. The relevant matrix element evaluates to 
\begin{align}
    \braket{2_p | e^{-iH_pt_2} | 2_p}  &=   \int dE \, \rho(E) e^{-iEt_2} \abs{ \braket{E | 2_p} }^2 \label{eqn:return_amplitude}\\
&+ e^{-iE_bt_2} \abs{ \braket{E_b | 2_p} }^2 \nonumber
\end{align}
where the energy $E$ is measured relative to the unperturbed two-planon state. The first term describes the continuum of extended states with $-8h_z\leq E\le8h_z$, while the second term accounts for a bound state at energy  $E_b<-8h_z$. For extended states in the absence of the hardcore constraint, the matrix element is approximately constant near the band edges, $\braket{E | 2} \sim 1/L$,  and the total density of states is also a constant and is extensive. The late-time asymptotics of the correlation function will be dominated by the band edges at $E \approx \pm 8\abs{h_z}$, giving rise to $\chi_e^{(1)}(t_2)\propto \sin({2\Delta^{(0)}_p t_2}) \, {\sin(8h_zt_2)}/{ t_2} $ (see Appendix \ref{Appendix_D} for derivation). The $1/t_2$ decay arises from the destructive interference of the continuum states.
When the hardcore constraint is included, the matrix element $\braket{E | 2_p}$ becomes energy-dependent at the band edges, which in turn modifies the late-time dynamics. In Appendix \ref{Appendix_D}, we show that the interaction leads to multiplicative polylogarithmic corrections to the linear response, viz.
\begin{equation}
    \chi_e^{(1)}(t_2) \propto \sin ({2\Delta^{(0)}_p t_2})
    \frac{ \sin(8h_zt_2)}{ t_2 \log(t_2)^2}
    .  \label{linear_response_extended}
\end{equation}
On the other hand, the bound state is separated from the continuum, so its linear response does not decay, $  \chi_b^{(1)}(t_2) \propto \sin ({(2\Delta^{(0)}_p +E_b) t} )$.
Combining these results with Eq.~\ref{eqn:pump_probe_scaling}, we obtain the envelope function of the pump-probe signal:
\begin{align}
\chi_\mathit{ZZX}(t_1,t_2)
&\propto
\begin{cases}
    \sqrt{t_2}/\log(t_2)^2, &\text{extended states}\\[2pt]
    t_2,   &\text{bound state}
\end{cases}
\end{align}
Observe that the late-time behavior is in fact dominated by the planon bound state.

We next discuss the effect of a realistic pump pulse on the braiding-induced nonlinear response. We consider a Gaussian pump pulse with duration $\tau$ and frequency detuning $\delta $ from the bottom of the two-lineon band (assuming $h\gg \delta, \tau^{-1}$), which excites a distribution of lineon pairs whose characteristic velocity and excitation density depend on the pump parameters. While these details modify the overall magnitude of the pump–probe signal, they do not affect its long-time scaling with $t_2$. In Appendix \ref{Appendix_F}, we show that the pump-probe coefficient (at a fixed time $t_2$) scales as  $\chi_\mathit{ZZX}(\tau, \delta) \propto \chi_Z^{(1)} \tau^0\delta^0 $ for $\delta\tau\ll 1$ and $
\chi_\mathit{ZZX}(\tau, \delta) \propto \chi_Z^{(1)} \tau\delta$ for $\delta\tau\gg 1$. We note that although the lineon density of states exhibits a van Hove singularity near the band edge, this singularity is regularized even at the level of linear response by the vanishing excitation matrix element. Nevertheless, the independence of $\chi_\mathit{ZZX}$ to $\delta \tau$ in the limit $\delta \tau \ll 1$ comes from the one-dimensional nature of the lineons. For braiding between two-dimensional particles, $\chi_\mathit{ZZX} \propto \sqrt{ \tau} $ in the limit $\delta \tau \ll1$ \cite{McGinley2024signatures} (whereas for fully three-dimensional point particles there is no braiding). Thus, the $\delta \tau \rightarrow 0$ limit of $\chi_\mathit{ZZX}$ itself contains a signature of lineons.

\begin{figure}[t]
\centering
\resizebox{0.8\columnwidth}{!}{%
\begin{tikzpicture}[
    line cap=round,
    line join=round,
    >={Stealth[length=4mm,width=2.6mm]}
]

\draw[black, line width=1.5pt]
  (-5.8+1-0.1,-2.1) -- (-1.9-0.1,1.92) -- (5.8-1-0.1,1.92) -- (1.9-1+1-0.1,-2.1) -- cycle;

  
\draw[orange!85!yellow, line width=2.2pt] (-2.6,0.33) -- (2.8,0.33);
\draw[blue,              line width=2.2pt] (-2.6,0.15) -- (0.5,0.15);
\draw[blue,              line width=2.2pt] (0.85,0.15) -- (2.8,0.15);

\draw[orange!85!yellow, line width=2.2pt, ->]
  (2,0.33) -- (1,0.33);
\draw[orange!85!yellow, line width=2.2pt, ->]
   (-2,0.33) -- (-1,0.33);

\draw[blue, line width=2.2pt, ->]
  (-0.85,0.15) -- (-1.95,0.15);
\draw[blue, line width=2.2pt, ->]
  (0.95,0.15) -- (1.95,0.15);

\draw[red, line width=2.2pt]
  (-3.55+0.2, -2.06)
    .. controls (-3.50+0.3,-1.35) and (-3.10+0.3,-1.10) .. (-2.30+0.3,-1.08+0.1)
    .. controls (-1.55,-1.06+0.18) and (-1.15,-1.12+0.22) .. (-0.75,-0.95+0.22)
    .. controls (-0.35,-0.7+0.1) and (-0.08,-0.30) .. (0.35,-0.10)
    .. controls (0.52,-0.02) and (0.63,0.06) .. (0.73,0.22);

\draw[red, line width=2.2pt]
  (0.80,0.46)
    .. controls (0.88,0.82) and (0.74,1.03) .. (0.55,1.18)
    .. controls (0.42,1.30) and (0.36,1.36) .. (0.38,1.56);

\draw[red, ->]  (0.74,1.03) -- (0.55,1.18) -- (0.42,1.30) ;
  
\node at (-2.85,0.25) {\Large$\ell$};
\node at ( 3.1,0.25) {\Large$\ell$};
\node at (0.65,1.55) {\Large$p$};

\node[blue] at (1.18,-0.18) {\Large 1};
\node[red] at (-0.2,-0.15) {\Large 2};
\node[orange!85!yellow] at (1.2,0.65) {\Large 3};

\end{tikzpicture}
}
    \caption{Sketch of a lineon pair braiding with a single planon in a two-dimensional spatial plane. One planon, located at the other end of the red trajectory, is not shown because the planon pair is far separated at late time $t_1$. The numbers denote the time ordering of the trajectories. Step 1: the lineon pair spreads out (blue). Step 2: the planon trajectory (red) crosses the lineon trajectory. Step 3: the lineons move back (orange) and are annihilated to generate the $\chi_\mathit{XXZ}$ signal. The orange and blue trajectories should lie on top of each other, but are slightly offset for visualization. This process generates a statistical phase of ($-1$).}
    \label{fig:Braiding_XXZ}
\end{figure}

\subsection{XXZ protocol}
Next, we discuss the XXZ  protocol which swaps the pump and probe pulses. In the XXZ protocol, a $z$-polarized pump pulse creates a pair of planons, while an $x$-polarized probe pulse applied at delay time $t_1$ creates a pair of lineons, and the resulting lineon-pair response is measured at time $t_1+t_2$. The nonlinear signal $\chi_\mathit{XXZ}$ measures the statistical interaction associated with a pair of lineons braiding with a single planon, as shown in Fig.\,\ref{fig:Braiding_XXZ}. In the corresponding spacetime diagram, the lineon pair forms a closed loop, which is intersected once by the planon's world-line, in analogy with Fig.\,\ref{fig:Braiding}.  Following the previous analysis, we treat the pumped planons (extended states) semiclassically with characteristic velocity $|\vec v|\propto\sqrt{h_z}$, while the probe lineons are treated quantum mechanically. The typical width of the lineons scales as $\Delta x_\perp \propto\sqrt{h_x t_2}$ in both the noninteracting and interacting cases (see Appendix D). As a result, the braiding probability and the effective braiding area satisfy
$
P[\vec r_{1,2}(t), |\vec v|]
\propto
A[\vec r_{1,2}(t); |\vec v|]
\approx
\Delta x_\parallel \Delta x_\perp
\propto
 |\vec v| t_2 \times \sqrt{h_x t_2},
$
which scales as $t_2^{3/2}$ as usual.
We next calculate the linear response for lineons. In the presence of the hardcore constraint, the matrix element between a nearby pair of lineons $\ket{2_\ell}$ and the eigenstates satisfies
$\braket{k|2_\ell} \propto \sin k \sim k$
at low energies. Consequently, the linear response for lineon pair is
\begin{align}
    \chi_X^{(1)}(t_2)
    &\propto  \braket{2_\ell|e^{-iH_\ell t_2}|2_\ell} \\
   &\propto\int_0^\infty dk \,
    e^{-i \varepsilon_\ell(k) t_2}
    |\braket{k|2_\ell}|^2
    \propto
    t_2^{-3/2}.
    \label{eqn:Lineon_linear_response}
\end{align}
where we disregard the overall oscillating phase factor and focus on the band edges, approximated as $\varepsilon_\ell(K=0,k) \approx \pm2h_x k^2$, with the center of mass quasimomentum set to zero in a translation-invariant system. We note that the hardcore constraint for one-dimensional fractionalized excitations significantly modifies the long-time behavior of the response. In the noninteracting case, $\braket{k|2_\ell}$ is approximately constant at low energies, leading to
$
\chi_X^{(1)}(t_2)\propto t_2^{-1/2}.
$
Combining the above results, the nonlinear XXZ signal approaches a constant magnitude at late times:
\begin{align}
      \chi_\mathit{XXZ}(t_1,t_2)
    &=
    (-2)\chi_X^{(1)}(t_2)
    P[\vec r_{1,2}(t), |\vec v|] \nonumber \\
    &\propto
    t_2^{-3/2} t_2^{3/2}
    =
    t_2^0.
\end{align}
In the XXZ protocol, the contribution from the planon bound state is negligible because the corresponding braiding probability $
P[\vec r_{1,2}(t), |\vec v|]\propto \sqrt{h_xt_2}
$ (only from the quantum spreading of the lineon pair in one dimension) grows parameterically slower than that of the extended states. A summary of the real-time linear, nonlinear ZZX and XXZ responses is presented in Table \ref{tab:response-scaling}. We emphasize that the late-time nonlinear XXZ signal is time-independent, in stark contrast to the linear-in-$t$ behavior of the ZZX signal. The asymmetry under reversing the polarizations is a direct consequence of the fact that the two species of fractionalized excitations (lineons and planons) live in different spatial dimensions, which is a characteristic signature of fracton phases. Furthermore, we can compare the enhancement ratio between the nonlinear and linear responses. In the XXZ protocol, $\chi_\mathit{XXZ}/\chi^{(1)}_X\propto t_2^{3/2}$, this scaling is claimed to be universal for the late-time braiding statistics in the 2D anyonic cases \cite{McGinley2024signatures}. In the ZZX protocol, we instead find $\chi_\mathit{ZZX}/\chi_Z^{(1)}\propto t_2$, which originates from the bound-state braiding statistics.


We next comment on the role of the static magnetic field $\vec h$, which 
serves as an additional parameter that can be tuned
in both ZZX and XXZ pump-probe experiment. First, since the excitation energies of lineon pairs and planon pairs depend on $h_x$ and $h_z$, respectively, varying $\vec h$ allows one to tune the resonance frequencies of the pump and probe pulses. Second, increasing $h_z$ enhances the magnitude of the planon bound-state signal relative to the extended state signal because the latter spreads out more rapidly ($\chi_e^{(1)}\propto 1/h_z$ in Eq.~\ref{linear_response_extended}). Moreover, increasing $h_z$ also increases the binding energy of the planon bound state, since the magnitude of the effective attractive potential is proportional to $h_z$ (Appendix \ref{Appendix_B}). As a result, the bound state becomes more robust against perturbations (e.g. short-range interactions) and is pushed farther away from the continuum in energy.
Third, tuning $\vec h $ also sets a crossover time scale for the nonlinear response of the extended states, due to the competition between the ballistic transport of the pump excitations and the quantum spreading of the probe excitations. To see this, consider the ZZX protocol and recall that the parallel width scales as $\Delta x_{\parallel}\sim |v|t_2+O(\sqrt{h_z t_2})$, where the subleading term arises from the quantum spreading of the planon pair. Since the lineon velocity scales as $|v|\propto \sqrt{h_x \delta}$ (for regime $\delta \tau \ll 1$), comparing the two contributions gives the critical time scale $t_c= O( \frac{h_z}{h_x\delta})$. For $t_2\gg t_c$, we recover the usual increasing signal
$
\chi_\mathit{ZZX}\propto \sqrt{t_2}/(\log t_2)^2
$ for the extended states,  whereas for $t_2\ll t_c$, the signal instead decreases with time as
$
\chi_\mathit{ZZX}\propto 1/(\log t_2)^2.
$ In the XXZ protocol, the critical time scale is instead given by
$
t_c= O(\frac{h_x}{h_z\delta}).
$ For $t_2\gg t_c$, we recover the late-time $t$-independent nonlinear signal, whereas for $t_2\ll t_c$, the nonlinear response decreases with time as
$
\chi_\mathit{XXZ}\propto t_2^{-1/2}.
$ Consequently, tuning the static magnetic field effectively determines whether the asymptotic late-time dynamics can be observed before the breakdown of perturbation theory as discussed in the next paragraph.

We have so far worked within a perturbative framework in which
braiding events involve only one of the fractionalized excitations of the pump (the time scales required to satisfy this criterion is discussed in Appendix F).
In this picture, the braiding probability increases with time indefinitely, implying an eventual breakdown of the perturbative expansion. This occurs because higher-order processes generate pump multiplets, and braiding events involving multiple particles (with even numbers giving a trivial phase) suppress the net statistical contribution, leading to a slowdown of the nonlinear signal. The time scale of the perturbative breakdown can be estimated as follows. Given a pump-induced excitation density $\rho$ and the effective braiding area $A[\vec r_{1,2}(t); v]$ computed before, we require the expected number of braided particles to satisfy $\braket {n}=\rho A[\vec r_{1,2}(t); v] \ll 1$. Consequently, for sufficiently low excitation density, there exists a long time window in which the perturbative regime is valid and the increasing pump-probe signal due to statistical interactions can be observed. In the ZZX protocol, since the effective braiding area of the planon's bound state grows as $ A[\vec r_{1,2}(t); v] \propto t_2$, which is slower than that of extended states $ A[\vec r_{1,2}(t); v] \propto t_2^{3/2}$, the increasing bound-state contribution remains perturbative to parametrically longer times.

%



\subsection{Scattering effects}

We have shown how short-range interactions, such as the hardcore constraints between identical fractionalized excitations, modify the linear and nonlinear response signals. We next consider short-range interactions between lineons and planons, in particular when the planons form a bound state. During a scattering event, the lineon can transfer (i) center-of-mass momentum $\vec K\neq 0$ to the planon bound pair along the lineon direction and (ii) relative momentum $\pm \vec k$ between the two planons. The latter process  corresponds to a transition from the bound state to the extended states where the transition probability can be estimated using the Fermi's golden rule. Since both the scattering potential and the bound-state wavefunction are short-ranged, while the low-energy extended states have suppressed weight near $\vec r=0$ due to the
hardcore constraint, the resulting matrix element is suppressed. Consequently,  the bound state signal is still dominant at late times, which continues to grow
linearly in $t$. However, scattering can also produce a phase shift of the quantum state, and therefore generate a nonzero pump-probe signal. The corresponding scattering probability is proportional to the length of the planon's worldline, giving a linear-in-$t$ enhancement to the pump-probe signal. Thus a linear-in-$t$
nonlinear signal is not, by itself, a unique fingerprint of braiding
statistics. Indeed, for a transverse-field Ising chain in the paramagnetic phase, the $q=0$ linear response for a single-quasiparticle excitation is non-decaying, while the scattering between quasiparticles created by the pump and probe can produce a linear-in-$t$ enhancement to the nonlinear signal \cite{Fava_2023}. Therefore, to more sharply identify braiding statistics and distinguish fractons from trivial phases, one can additionally perform the XXZ experiment. In this protocol, the scattering probability grows parametrically slower than the braiding probability, so that braiding statistics dominates at late times. Specifically for the lineon-pair responses, we obtain $\chi_\mathit{XXZ}/\chi_X^{(1)}\propto t^{3/2}$ and $\chi_\mathit{XXZ}\propto t^0$. This asymmetry under swapping pump and probe polarizations would be absent in trivial phases, and we believe is a fingerprint of fracton phases.

\section{Conclusion}
We have demonstrated how the existence of nontrivial lineon-planon braiding statistics can be probed spectroscopically in the X-cube model. Our discussion builds on analogous treatments for spin liquids~\cite{McGinley2024signatures,Yang_PNAS,kirchner2025measuring}, but has important differences. In particular, the relevant pump-probe signal $\chi_\mathit{ZZX}$ exhibits asymptotic linear-in-$t$ behavior (with $\chi_Z^{(1)}\propto t^0$) that differs from Refs.~\cite{McGinley2024signatures,Yang_PNAS,kirchner2025measuring}, and is dominated by planon bound states, which have no analog in conventional spin liquids. The emergent bound state originates from the ability of nearby planons to explore different planes through alternative pairing configurations. Furthermore, we demonstrate a strong asymmetry under swapping the pump and probe polarizations, where the nonlinear response $\chi_\mathit{XXZ}$ instead exhibits asymptotically $t$-independent behavior (with $\chi_X^{(1)}\propto t^{-3/2}$), which provides evidence that the two species of fractionalized excitations live in different spatial dimensions.
{In this work, the pump-probe signals (i) demonstrates the existence of non-trivial braiding statistics between excitations in a three dimensional system, and (ii) is sensitive to the existence of a bound state between some of the excitations with nontrivial braiding statistics, and (iii) to some of the fractionalized excitations being mobile in only one dimension. The combination of these features is highly suggestive of fractionalized excitations with subdimensional mobility, as is characteristic of (type-I) fracton phases.}
The discussion illustrates how pump-probe techniques can be used to formulate highly specific diagnostics for fracton phases, and how these diagnostics can differ in fracton phases from how they work in more traditional spin liquids. We believe that many key qualitative features studied in this work persist beyond the X-cube model. In particular, the asymmetry under reversing the pump and probe polarizations, which originates from spatial asymmetry of the subdimensional excitations, is expected to be a robust feature shared by many type-I fracton models containing both lineons and planons. In addition, we expect that the mechanism underlying the planon bound state can extend to a broad class of type-I fracton phases, although a careful treatment of the hardcore constraint is required in models with different geometries. An extension of these ideas to other fracton phases is a worthwhile problem that is left for future work.

{\bf Acknowledgements}: We acknowledge helpful discussions with Evan Wickenden. This work was supported by  the U.S.\ Department of Energy, Office of Science, Basic Energy Sciences under Award No.\ DE-SC0021346.
\appendix

\section{Boundary conditions of lineons \label{Appendix_A}}
We consider a pair of lineons in a finite-size system, where the separation between the two lineons satisfies $1 \leq x \leq L-1$. For convenience, we will work with system sizes with $L$ odd.
Note that (i) if $R$ is an integer $x$ must be even, while if $R$ is a half-odd-integer then $x$ must be odd, and (ii) in order to correctly handle periodic boundary conditions we must identify the states $\ket{R, x} \equiv \ket{R+L/2, L-x}$.
We therefore choose to work with states $\ket{R, x}$ such that $x < L/2$ (while $R$ is unconstrained), all of which are independent. 

Acting with~\eqref{eqn:X-cube-perturbed} on the lineon-pair state $\ket{R, x}$, we have
\begin{align}
     H_\text{XC}(h_x, h_z) \ket{ R, x }= &-h_x [\ket{R-\tfrac12, x+1} +\ket{R-\tfrac12, x-1}  \nonumber \\
     &+\ket{R+\tfrac12, x+1} +\ket{R+\tfrac12, x-1} ]
\end{align}
in the bulk. 
The four states on the right correspond to each of the two lineons moving to the left and the right. 
Higher-order expressions in $h_x$, $h_z$ lead to hopping over larger distances.
When $x=1$, this is modified to
%
%
\begin{equation}
    H_\text{XC}(h_x, h_z) \ket{ R, x } = -h_x\ket{R-\tfrac12, x+1} -h_x\ket{R+\tfrac12, x+1}
    ,
\end{equation}
%
%
since the lineons cannot be separated by less than one lattice site. 
For $L$ odd and $x=\lfloor L/2 \rfloor$ we have
\begin{align}
     H_\text{XC}(h_x, h_z) \ket{ R, x } = &-h_x [\ket{R-\tfrac12, x-1} +\ket{R+\tfrac12, x-1} \nonumber \\
     &+\ket{R+\tfrac{L}{2}+\tfrac12, x}  +\ket{R+\tfrac{L}{2}-\tfrac12, x}] 
     ,
\end{align}
since we have chosen to work with states satisfying $x < L/2$.
The eigenstates and energies are found by Fourier transformation. In particular, let $\braket{ R, x | \Psi} = \Psi(R, x) = A e^{iKR}\sin kx$. Then $H\ket{\Psi} = E\ket{\Psi}$ with
\begin{equation}
    E(K, k) = - 4h_x\cos(K/2)\cos k = -4h_x + \varepsilon_l(K, k)
    .
\end{equation}
Therefore, the energy cost to create a pair of lineon excitations is modified to $2\Delta_l = 8J_z - 4h_x$.
The wave vectors $k$ and $K$ are quantized according to the solutions of
\begin{align}
    \frac{\sin[k(\lfloor L/2 \rfloor - 1)]}{\sin[k\lfloor L/2 \rfloor]} = 2\cos k - e^{iKL/2} \\
    \implies 
    k = 
    \begin{cases}
        \frac{\pi(2n_+ + 1)}{L} &\text{ if } e^{iKL/2}=1 \\
        \frac{2\pi n_-}{L}      &\text{ if } e^{iKL/2} = -1
    \end{cases},
\end{align}
and $K = 2\pi m / L$, respectively. Note that $e^{iKL/2} \in \{1, -1 \}$. Linearly independent for $m \in \{0, 1, \dots , L-1\}$, $n_+ \in \{ 0, 1, \dots , \lfloor L/2 \rfloor - 1 \}$, $n_- \in \{ 1, 2, \dots \lfloor L/2 \rfloor \}$, giving $L \lfloor L/2 \rfloor = L(L-1)/2$ states in total, as required. The normalization constant is $A = 2/L$.
Note that this analysis accounts for both the identical nature of the lineons and the hardcore constraint $x \geq 1$ \emph{exactly} to first order in the magnetic field strength $h_x$.
\section{Planon bound state \label{Appendix_B}}

A pair of planons under static magnetic field can be mapped to an effective 2D tight-binding model (Eq.~\ref{eqn:planon_hamiltonian}). Focusing on the uniformly excited center-of-mass sector with quasimomentum $\vec K=0$, the two-body Hamiltonian in Eq.~\ref{eqn:planon_hamiltonian} reduces to a single-particle problem for the relative coordinate $\vec r$:
\begin{align}
H_{p} = &-2h_z \sum_{\vec{r}} \sum_{\vec\delta = \pm \hat{x},\,  \pm \hat{y}}
 \lvert  \vec{r} + \vec\delta \rangle \langle \vec{r} \rvert  
+ \sum_{\vec{r}} V(\vec{r}) \lvert   \vec{r} \rangle \langle \vec{r} \rvert ,
\label{eqn:planon_hamiltonian_r}
\end{align}
Note that the effective hopping becomes $2h_z$ since either planon can hop and thus changes the relative coordinate with the same amplitude. The band dispersion is $E(\vec k )=-4h_z(\cos k_x+\cos k_y)$. The local potentials are

\begin{equation}
V(\vec{r}) =
\begin{cases}
\lambda\rightarrow\infty, & \vec{r} = (0,0), \\
U(E), & \vec{r} = (\pm1,0);(0,\pm1) \\
0, & \text{otherwise}.
\end{cases}
\end{equation}
where the energy dependent local potentials  $U(E)$ originate from projecting out the extra degrees of freedom. Below the two-planon continuum $E<-8h_z$, the potential is attractive with $U(E)<0$. In the absence of the hardcore constraint, any arbitrarily weak attractive potential produces at least one bound state. However, the hardcore constraint acts as an infinite repulsive potential at the origin, which introduces a threshold  $|U|>U_c$ for the bound-state solution. In the following, we first derive the threshold $U_c$ in the presence of hardcore potential, then we calculate the effective energy-dependent potential $U(E)$ obtained by projecting out the extra degrees of freedom, and show that  $|U(E)|_\text{max}>U_c$ satisfies the bound-state criterion for $E<-8h_z$. This proves the existence of a planon bound state and allows us to calculate the bound state energy. 


We first decompose the perturbation into a repulsive part $V_1=\lambda \ket{0}\bra{0}$ and an attractive part $V_2 =U\sum_{\vec{r}_i}\ket{\vec{r}_i}\bra{\vec{r}_i}$. 
We begin by computing the Green function under the perturbation $V_1$. The Dyson equation reads 
\begin{equation}
    G=G^0+G^0V_1G^0+G^0V_1G^0V_1G^0+...=G^0+G^0V_1G.
\end{equation}
where $G^0$ is the Green function for the unperturbed system. In real space this gives  $G_{\vec{r},\vec{r}'}=G^0_{\vec{r},\vec{r}'}+\lambda G^0_{\vec{r},\vec{0}} G_{\vec{0},\vec{r}'}$. Solving for $ G_{\vec{0},\vec{r}'}$ and taking $\lambda\rightarrow\infty$, we find: 
\begin{align}
    G_{\vec{r},\vec{r}'}(E)&=
    G^0_{\vec{r},\vec{r}'}+\lambda \frac{ G^0_{\vec{r},\vec{0}} G^0_{\vec{0},\vec{r}'}}{1-\lambda G^0_{0,0}}
    =G^0_{\vec{r},\vec{r}'}- \frac{ G^0_{\vec{r},\vec{0}} G^0_{\vec{0},\vec{r}'}}{ G^0_{0,0}} \nonumber \\
    &= G^0_{\vec{r}-\vec{r}'}- \frac{ G^0_{\vec{r}} G^0_{\vec{r}'}}{ G^0_{0}}
    \label{eqn:perturb_G}
\end{align}
where in the last step we used translational invariance of the unperturbed system. This relation connects the Green function in the presence of the hardcore constraint to the Green function of the clean system. 

We now introduce the attractive potential 
\begin{equation}
    V_2 =U\sum_{\vec{r}_i}\ket{\vec{r}_i}\bra{\vec{r}_i} =UP
\end{equation}
where $U<0$ and $P$ is the projector onto the subspace spanned by $\ket{\vec{r_i}}$ with $\vec{r}_1 = (1,0), \vec{r}_2 = (-1,0), \vec{r}_3 = (0,1), \vec{r}_4 =(0,-1)$. The full Green function $\widetilde G$ becomes: 
\begin{equation}
    \widetilde G = \frac{G}{1-V_2G}.
\end{equation}
where $G$ is the Green function with a hardcore constraint at the origin (with $V_1$ perturbation). The Bound states correspond to poles of  the full Green function, that is $\text{det}(1-V_2G)=0$. Dividing the Hilbert space into the subspace spanned by $\ket{\vec{r}_i}$ with projector $P$ ($P^2=P$) and its complement (with projector $1-P$), the matrix can be written in block form as:
\begin{equation}
    1-V_2G = 
\begin{pmatrix}
 1-UPGP &  -UPG(1-P)    \\
 0     &  1 
\end{pmatrix}.
\end{equation}
The bound state solutions becomes
\begin{align}
\text{det}(1-V_2G) &= \text{det}(1-UPGP) 
\\
&=\text{det}(\delta_{ij}-UG_{\vec{r_i},\vec{r_j}}(E))\\
&=0
\end{align}
where $G_{\vec{r_i},\vec{r_j}} = \braket{\vec r_i|G|\vec r_j}$. Therefore, the bound state solutions can be evaluated entirely within the subspace formed by $\ket {\vec r_i}$. 
For simplicity we define $A\equiv G_{\vec{r}_1,\vec{r}_1}$, $B\equiv G_{\vec{r}_1,\vec{r}_2}$ and $C \equiv G_{\vec{r}_1,\vec{r}_3}$. Due to symmetry, the $4{\times}4$ matrix in the subspace can be explicitly written as
\begin{equation}
    1-UPGP = 1- U
\begin{pmatrix}
 A &  B     &  C     &  C \\
 B     &   A &  C     &  C \\
 C     &  C     &  A & B \\
 C     & C     &  B     &  A
\end{pmatrix}.
\end{equation}
Next, we can change the basis by defining the $s, p_x, p_y, d$ states as linear combinations of $\ket{\vec{r}_i}$:
\begin{align}
    \ket{s}&=(1,1,1,1)^T/2\\\
    \ket{p_x}&=(1,-1,0,0)^T/\sqrt{2}  \\
    \ket{p_y}&=(0,0,1,-1)^T/\sqrt{2} \\
    \ket{d}&=(1,1,-1,-1)^T/2 
\end{align}
Under the change of basis, the matrix is diagonalized as $\text{diag}(1-Ug_s, 1-Ug_{p_x}, 1-Ug_{p_y}, 1-Ug_{d}  )$ where $g_\alpha =\braket{\alpha|G(E)|\alpha}$. We note that only the s and d-wave bound states are allowed by symmetry ($\vec{r} \rightarrow-\vec{r}$).  Therefore, the bound state conditions (the poles of the Green function) become:
\begin{align}
       U  &= 1/g_s(E)  \\
       U  &= 1/g_d(E)  
\end{align}
It can be shown that $g_s(E) = A+B+2C$ while $g_d(E)=A+B-2C$. Using the relation between $G$ and the unperturbed Green function $G^0$ in equation \ref{eqn:perturb_G}, we obtain $A=G^0_{0}-(G^0_{(1,0)})^2/G^0_{0}$, $B=G^0_{(2,0)}-(G^0_{(1,0)})^2/G^0_{0}$ and $C=G^0_{(1,1)}-(G^0_{(1,0)})^2/G^0_{0}$. As a result,
\begin{align}
    g_s(E) &=G^0_{0} +G^0_{(2,0)}+ 2G^0_{(1,1)} - 4 \left(G^0_{(1,0)}\right)^2/G^0_{0} \\
    g_d(E) &=G^0_{0} + G^0_{(2,0)} - 2G^0_{(1,1)}
\end{align}
where
\begin{align}
    G^0_\mathbf r(E)
&= \bigl\langle \mathbf r \bigl| \frac{1}{E + i0^{+} - H_0} \bigr| \mathbf 0 \bigr\rangle \nonumber
\\&= \frac{1}{(2\pi)^2}\int d^2\vec k\;
\frac{e^{i \vec{k}\cdot \vec{r}}}
     {E +i0^+-E(\vec k)}.
\end{align}
where $E(\vec k)=-4h_z\bigl(\cos k_x + \cos k_y\bigr)$. Below the continuum, $E<-8h_z$, the Green functions $G^0_\mathbf r(E)<0$ and $g_\alpha(E)<0$ are negative. 
The clean Green function $G^0_\mathbf r(E)$  vanishes as $E\rightarrow-\infty$, decreases monotonically as $E$ increases, and diverges to $-\infty$ as $E\rightarrow-8h_z^-$. Thus, in the absence of the hardcore constraint, any small attractive potential with $U<0$ can satisfy the bound state condition. However, with the hardcore constraint, $g_\alpha(E)$ no longer diverges at the band edge because the divergence of the positive and negative terms cancel. Instead, it approaches a finite value $g_{\min}$ as $E\rightarrow-8h_z^-$, leading to a threshold for bound-state formation for channel $\alpha$: 
\begin{equation}
    |U|\geq  |g_{\alpha,\text{min}}|^{-1}\equiv U^{(\alpha)}_c
\end{equation}
Numerical results for $|g_{s}(E)|^{-1}$ and $|g_{d}(E)|^{-1}$ are shown in Fig.~\ref{fig:bound_state} where $|g_{s,\text{min}}|^{-1} = 2h_z$ and  $|g_{d,\text{min}}|^{-1} \approx 7.32h_z$ occurs at the band edge. For comparison, we also plot $|g^0_s(E)|^{-1} $ where $ g^0_s(E)=G^0_{0,0} +G^0_{2,0}+ 2G^0_{1,1}$ corresponds to the $s$ wave channel in the absence of the hardcore constraint, which exhibits the expected divergence $|g^0_s(E)|^{-1}\rightarrow0$ at the band edge.

\begin{figure}[t]
    \centering
    \includegraphics[width=0.48\textwidth]{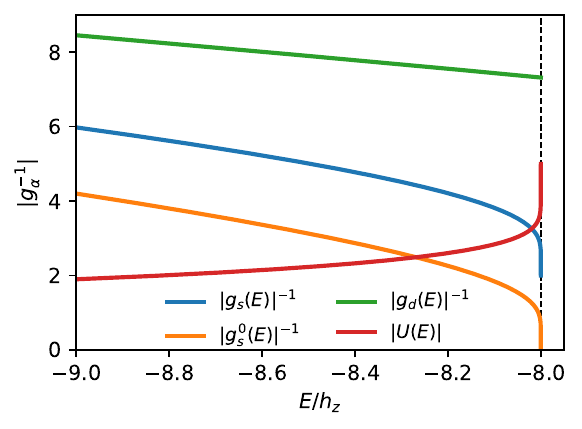}
        \caption{The inverse Green's functions $|g_s(E)|^{-1}$ and  $|g_d(E)|^{-1}$ denote the $s$- and $d$- wave channel in the presence of hardcore constraint, while $|g^0_s(E)|^{-1}$ is the $s$-wave Green's function without the hardcore constraint. The edge of the two-planon continuum locates at $E/h_z =-8$. The red curve shows the energy-dependent effective attractive potential $|U(E)|$ obtained by integrating out the extra degrees of freedom. The bound state solution is given as $|U(E)|=|g_\alpha(E)|^{-1}$. }
    \label{fig:bound_state}
\end{figure}

Our next step is to calculate the effective potential $U(E)$ below the planon continuum. The attractive potential arises from projecting out the extra degrees of freedom. In the two-dimensional description, the points $\vec{r}_{xy}=(1,0)$ and $\vec{r}_{yz}=(0,1)$ connect the $xy$ and $yz$ planes, but they represent the same state in the full Hilbert space. We must therefore remove this redundant degree of freedom. We choose to keep $\vec{r}_{xy}=(1,0)$ in $H_{xy}$ while excluding the point $\vec{r}_{yz}=(0,1)$ from $H_{yz}$. From this perspective, excluding $\vec{r}_{yz}=(0,1)$ is equivalent to imposing an infinite repulsive potential at that site. As a result, the coupling between the two planes can be written as $\widetilde V = -2h_z \sum_\vec{r_i}\ket{1,0}_{xy}\bra{\vec{r_i}}_{yz}$ where
$\ket{\vec{r_i}}_{yz} = \ket{1,1}, \ket{-1,1},\ket{0,2}$ live in the $yz$-plane. The effective Hamiltonian obtained after integrating out the $yz$-plane is
\begin{align}
    H_\text{eff} &= H_{xy} +\widetilde{V} \frac{1}{E-H_{yz}}\widetilde V^\dagger \nonumber\\
    & = H_{xy} + 4 h_z^2\sum_\vec{r_i, r_j} \braket{\vec{r_i}|\frac{1}{E-H_{yz}}|\vec{r_j}}  \ket{1,0}\bra{1,0}_{xy}\nonumber \\
    &  = H_{xy} +U(E)  \ket{1,0}\bra{1,0}_{xy} 
\end{align}
Thus the effective on-site potential at $\vec{r}_{xy}=(1,0)$ is given by summing over the Green functions of the $yz$-plane with the site $\vec{r}_{yz}=(0,1)$ removed. Shifting the coordinates in the $yz$-plane such that the infinite potential at $\vec{r}_{yz}=(0,1)$ now sits at the origin and using the relation in Eq.~\ref{eqn:perturb_G}, we obtain:
\begin{align}
     U(E) &= 4h_z^2\sum_\vec{r_i, r_j} G_\vec{r_i, r_j}(E)  \label{eqn:attractive_potential} \\
     &= 4h_z^2 \left(3G^0_{0}+4G^0_{(1,1)}+2G^0_{(2,0)}- 9\left(G^0_{(1,0)}\right)^2/G^0_{0} \right) \nonumber \\
     &<0 \nonumber
\end{align}
at location $\vec{r}_{xy}=(1,0)$. Since we identify $\vec{r}$ and $ -\vec{r}$, the same local potential should also exist at $\vec{r}_{xy}=(-1,0)$. Similarly, projecting out the $xz$-plane would lead to the same local potentials at $\vec{r}_{xy}=(0,\pm1)$. 

The magnitude of the local potential $|U(E)|$ is energy dependent and increases monotonically as $E$ approaches the band edge from below. In the limit $E\rightarrow-8h_z^-$, we find $|U|_\text{max} = 5h_z$. Comparing this with the thresholds obtained above, we observe that $|U|_\text{max}>U^{(s)}_c=2h_z$ for the $s$-wave channel, whereas $|U|_\text{max}<U^{(d)}_c\approx 7.32h_z$ for the $d$-wave channel. Consequently, only a single s-wave planon bound state appears in the energy gap. In Fig.~\ref{fig:bound_state}, we plot $|U(E)|$ along with $|g_\alpha(E)|^{-1}$, suggesting a $s$- wave bound state solution at a crossover as $|U(E)|=|g_s(E)|^{-1}$. 

Inside the two-planon continuum $-8h_z\leq E\leq8h_z$, the effective local potential $U(E)$ becomes complex. The real part is of order $h_z$, and changes sign between $E<0$ and $E>0$. Recall that in the full interacting potential $V(r)$, the hardcore constraint at the origin has magnitude $\lambda=4J_x \gg h_z$. As a result,  the real part of $U(E)$ on sites surrounding the origin can be viewed as a small renormalization of the effective hardcore strength and radius.  In contrast, the imaginary part of $U(E)$ encodes the non-Hermitian nature of the effective two-dimensional planon-pair problem since the wavefuncation can leak out to other 2D planes. Since the late-time dynamics is dominated by low energies, we focus on the unperturbed Green function near the bottom of the planon band. Approximating the dispersion by $E(k) \approx -8h_z+2h_z k^2$, the imaginary part of the unperturbed Green's function at energy $E=-8h_z+\epsilon$ can be evaluated as
\begin{align}
    \text{Im} \,G^0_\vec r( E) &= \frac{1}{(2\pi)^2 } \int d^2 \vec k \; e^{i\vec k \cdot \vec r} \left[ -\pi \delta(E-E(\vec k)\right)] \nonumber \\ 
   & = -\frac{1}{4\pi}  \int^\infty _0 dk \,k \, \delta(\epsilon -  2h_z k^2) \int^{2\pi}_0 d\theta\, e^{ikr\cos\theta}  \nonumber \\
   & = -\frac{1}{8h_z} J_0\left(\sqrt{\frac{\epsilon}{2h_z}}r\right) \nonumber \\
   &\approx -\frac{1}{8h_z} \left( 1-\frac{\epsilon}{8h_z}r^2 \right)
    \end{align}
where $J_0$ is the Bessel function of the first kind. As $\epsilon \rightarrow0^+$,  $\text{Im} \,G^0_\vec r( E)$ approach a constant $-1/8h_z$ . In the same limit, $\text{Im}\,U(E)$ in Eq. \ref{eqn:attractive_potential} vanishes,  so the effective imaginary potential becomes relatively small near the band edge. Therefore, we conclude that  $\text{Im}\,U(E)$ has a negligible effect on the late-time dynamics of interest.

\section{Linear response for planons \label{Appendix_D}}

We consider the effective Hamiltonian for the planon pair in Eq.~\ref{eqn:planon_hamiltonian_r} and impose {open boundary conditions} on the relative coordinate. Having ignored the interacting potential $V(\vec r)$,  the solutions are
\begin{align}
     C_\vec{k}(\vec{r}) &= \frac{1}{N+1} \cos(k_x x) \cos(k_y y) \\
    S_\vec{k}(\vec{r}) &= \frac{1}{N+1} \sin(k_x x) \sin(k_y y),
\end{align}
which are symmetric under $\vec{r} \to - \vec{r}$. Here, the relative coordinate is defined on $x,y \in [-N,N]$ where $N= \lfloor L/2 \rfloor$. In $C_\vec{k}$, the quasimomentum $\vec{k}$ is quantized as $k_\alpha = (n_\alpha+1/2)\pi/(N + 1)$ for $n_\alpha \in \{ 0, 1, \dots, N\}$. In $S_\vec{k}$, it is quantized as $k_\alpha = n_\alpha\pi / (N + 1)$ for $n_\alpha \in \{ 1, \dots, N \}$, giving $2N^2 + 2N + 1$ states.
This is consistent with the expected total number of states, equal to $[(2N + 1)^2-1]/2+1$.

The quantity of interest is the return amplitude in Eq.~\ref{eqn:return_amplitude}. Here, we focus only on the extended states contribution where late times are dominated by the low-energy behavior of the overlap $\braket{E | 2_p}$ with the initial state $\ket{2_p} = \frac{1}{2}(\ket{(1,0)}+\ket{(-1,0)}+\ket{(0,1)}+\ket{(0,-1)})$. The overlap between $\ket{2_p}$ and $\ket{C_\vec{k}}$
is equal to $ (\cos k_x + \cos k_y)/(N+1) \approx 2/(N+1)$, while its overlap with $\ket{S_\vec{k}}$ vanishes. At low energies, the density of states is also independent of energy, behaving as $\Theta(E)$ up to constants. Including only the states $C_\vec{k}$, the density of states that contribute to the late-time behavior is equal to $(N+1)^2 /8\pi h_z$. Note that, on the lattice, we must also account for the high-energy states at $E \approx 8\abs{h_z}$ where the density of states is again constant near the top of the band. The late-time asymptotics of the correlation function will therefore be dominated by both band edges $E \approx \pm 8{h_z}$. We can then evaluate the late-time behavior of the return amplitude for the planon extended state:
\begin{align}
    I(t) &=  \braket{2_p | e^{-iH_pt_2} | 2_p} \\
    &= \int dE \, e^{-iEt} \rho(E) \abs{ \braket{E | 2_p} }^2 \\
    &= \frac{i}{t}\left[ e^{-iEt} \rho(E) \abs{ \braket{E | 2_p} }^2 \right]_{-8h_z}^{+8h_z} + O(t^{-2}) \\
    &\approx \frac{i}{2\pi h_z t} \left[ e^{-i8h_zt} - e^{i8h_zt} \right] \\
    &=\frac{1}{\pi h_z t} \sin(8h_zt).
\end{align}
%
%
which has a $1/t$ decay envelope arising from the long-wavelength behavior, and it oscillates at a frequency set by the lattice bandwidth. The linear response is then calculated by
\begin{align}
 L^3 \chi_Z^{(1)}(t_2) &= -i\Theta(t) \braket{[A(t_1+t_2),A(t_1)]}_0 \\
                          &=2\text{Im} \braket{A(t_1 + t_2) A(t_1)}_0
\end{align}
where we use the fact that $A$ is Hermition. Using Eq. \ref{eqn:correlation}, we obtain:
\begin{equation}
    \chi_Z^{(1)}(t_2) = \frac{2\sin(2\Delta_p^{(0)}t_2)\sin(8h_zt_2)}{\pi h_z t_2}.
\end{equation}

In the full interacting problem, there are additional effective local potentials surrounding the origin. For the band eigenenergies, the energy shifts induced by such local perturbation scale as $O(1/L^2)$. Consequently, the densities of states and the energy spectrum of the continuum remain unchanged in the thermodynamic limit. However, the eigenstates would acquire phase shifts due to the repulsive potential which acts as a scatterer. This leads to an energy dependent of the overlap $\braket{E|2_p}$ in low-energies. To make further analytical progress, we write lattice Laplacian in terms of a continuum derivative. This approximation is justified when the wavefunction varies smoothly on the lattice scale. The equation we would like to solve is
\begin{equation}
    [- \nabla^2 + V_{\text{hc}}(r)] \psi(r, \theta) = \varepsilon_p \psi(r, \theta)
\end{equation}
where $V_{\text{hc}}(r)$ is taken to be infinite for $r\leq b$ and zero otherwise, which encodes the hardcore constraint over a length scale $b$. The energy $\varepsilon_p= E + 8h_z \approx 2h_zk^2$ is the energy relative to the bottom of the band. In cylindrical coordinates, this becomes
\begin{equation}
    -\frac{\partial^2 \psi}{\partial r^2} - \frac{1}{r} \frac{\partial \psi}{\partial r} - \frac{1}{r^2} \frac{\partial^2 \psi}{\partial \theta^2} + V_{\text{hc}}(r) \psi = \varepsilon_p \psi
\end{equation}
For simplicity, we modify the IR boundary conditions to be cylindrically symmetric. Namely we impose open boundary conditions, $\psi(R, \theta)=0$ at $r=R$. Since the problem now has cylindrical symmetry, angular momentum is a good quantum number and the general solution for the relative-coordinate wavefunction is $\psi = e^{il\theta} \phi_l(r)$, where $\phi_l(r)$ satisfies the radial equation
\begin{equation}
    -\frac{d^2 \phi_l}{d r^2} - \frac{1}{r} \frac{d \phi_l}{d r} + \frac{l^2}{r^2} \phi_l + V(r) \phi_l = \varepsilon_p \phi_l
    .
\end{equation}
Since the planons are bosonic, the angular momentum quantum number $l \in 2\mathbb{Z}$, giving $\psi(r, \theta) = \psi(r, \theta+\pi)$.
For $r < b$, we have $\phi_l(r) = 0$ since $V(r)$ is infinite. In the region $r > b$, the general solution to this equation is
\begin{equation}
    \phi_{l} (r) = A_l J_{\abs{l}} (kr) + B_l Y_{\abs{l}} (kr)
\end{equation}
where $J_l$ and $Y_l$ are Bessel functions of the first and second kind, respectively. We now implement the boundary conditions at $r=b$ and $r=R$. The vanishing of the wave function at $r=b$ requires that 
\begin{equation}
    \tan \delta_l = -B_l / A_l = J_{\abs{l}}(kb) / Y_{\abs{l}}(kb),
\end{equation}
while the discreteness of the spectrum derives from the IR boundary condition, $\phi_l (R)=0$. Up to normalization, the full solution (including COM coordinate $\vec R$) is therefore
\begin{equation}
    \Psi(\vec{R}, r, \theta) = C e^{i \vec{K} \cdot \vec{R}} e^{i l \theta} [ J_{\abs{l}}(kr) - \tan\delta_l Y_{\abs{l}}(kr) ]
\end{equation}
where $\delta_l$ depends implicitly on the hardcore radius $b$ and the relative momentum $k$. In the following, we set COM quasimomentum $\vec K = 0$.  


In the interacting case, the late-time behavior is dominated by the $l=0$ channel. In this case,
\begin{equation}
    \langle E | 2_p \rangle = 2 C_0(k)\left[ J_{0}(k) - \tan\delta_0 Y_{0}(k) \right]
\end{equation}
where $C_0(k)$ is a normalization constant.
Explicitly, the normalization constant is found by evaluating 
\begin{align}
        &\int_b^R 2\pi r dr \, [J_{0}(kr) - \tan\delta_0 Y_{0}(kr)]^2 \nonumber \\
        &= \pi \left( R^2 \varphi_{1,k}^2(kR) - b^2 \varphi_{1,k}^2(kb) \right)
    \label{eqn:hardcore-normalization}
\end{align}
where $\varphi_{l,k}(kr) = J_{\abs{l}}(kr) - \tan\delta_l(k) Y_{\abs{l}}(kr)$ is the unnormalized wavefunction. Note that in Eq.~\eqref{eqn:hardcore-normalization} $k$ is implicitly determined by the quantization condition $\varphi_{0,k}(kR) = 0$. Thus, $\varphi_{1,k}$ is being evaluated at $k$ corresponding to $l=0$ and $\varphi_{1,k}(kR)$ does \emph{not} vanish. In the long-wavelength limit, $kb \ll 1$, we have $\tan\delta_0 \sim (\log kb)^{-1} \ll 1$. Note that system size $R$ must be \emph{very} large in order for $\log b/R$ to be negligible. Nevertheless, in the remainder of this derivation we will assume that $\log b/R \ll 1$. In this case, the quantization condition is approximately $J_{0}(kR) = 0$ and the allowed values of $kR$ are determined by the Bessel function zeros $j_{0,m}$ with the asymptotic expansion $j_{l,m} = (m+l/2-1/4)\pi + O(1/(m+l/2))$. Therefore, for $R \gg b$ the normalization is approximately
\begin{align}
    &\int_b^R 2\pi r dr \, [J_{0}(kr) - \tan\delta_l Y_{0}(kr)]^2  \approx \pi R^2 J_1^2(kR) \nonumber \\
   & \approx \frac{2R}{k} + O[(k\log k)^{-1}] 
\end{align}
where we also assumed that $kR \gg 1$, equivalent to $m \gg 1$ in the Bessel function zeros $j_{0,m}$. Hence, we obtain $C_0(k) \approx \sqrt{k/2R}$. 

Near the bottom of the band, we can approximate $\varepsilon_p=E+8h_z\approx 2h_zk^2$. 
In the long-wavelength regime $k\ll1$, the Bessel functions can be approximated by
\begin{equation}
    J_0(x) \approx 1, \quad 
    Y_0(x) \approx \frac{2}{\pi} \left[ \log\frac{1}{2}x + \gamma \right].
\end{equation}
The logarithmic singularity of $Y_0(x)$ at $k=0$ will dominate the late-time behavior.
At low energies, the matrix element is approximated by 
%
%
\begin{align}
    \braket{E | 2_p} &\approx 2C_0(k) \left[ 1 - \frac{\log (k/2) + \gamma}{\log (bk/2) + \gamma} \right] \nonumber \\
    &=  \frac{2C_0(k)\log b }{\log k + \log b/2 + \gamma} \nonumber \\
   & \approx  \sqrt\frac{{2k}}{R}\frac{\log b }{\log k} 
\end{align}
with an analogous expression at the top of the band near $E \approx 8 h_z$. Next, we evaluate the low-energy density of states for the $l = 0$ states $\rho(E) = \frac{dN}{dk} \frac{dk}{dE} \approx \rho_0/(4h_z k) $ where $\rho_0 \equiv  dN/dk$ can be obtained by imposing the boundary condition:
\begin{equation}
    J_{0} (kR) -\tan \delta_0 Y_{0} (kR) \stackrel{!}{=} 0.
\end{equation}
At low energies, the arguments $kR \sim 1$, implying that $kb \ll 1$. Therefore $Y_{0} (kR) \ll Y_{0} (kb)$ and the solutions are given approximately by the solutions of
\begin{equation}
    J_{0} (kR) = 0
    ,
\end{equation}
whose density is $\rho_0 = R/\pi$. We therefore arrive at the final result for the return amplitude:
\begin{align}
    I(t) &= \int dE \, e^{-iEt} \rho(E) \abs{ \braket{E | 2_p} }^2 \nonumber \\
    &= \frac{2}{R} (\log b)^2 \rho_0 \left[ e^{-8ih_zt} - e^{8i h_z t} \right] \frac{i}{h_z t \log(t)^2} \nonumber \\
    &=  \frac{4 (\log b)^2 \sin(8h_zt) }{\pi h_z t \log(t)^2} 
    ,
\end{align}
Combining it with the fast oscillation $\sin(2\Delta^{(0)}_pt)$ from the two-planon energy gap, we obtain the linear response signal $\chi^{(1)}_Z(t)$ as shown in Eq.\,\ref{linear_response_extended}. This result shows that the interactions lead to polylog corrections in time to the noninteracting $1/t$ decay.

\section{Planons' and lineons' typical width \label{Appendix_E}}
Now, we derive the typical separation of a planon pair that returns to the initial state $\ket{2_p}$ after a time interval $t_2$. The typical separation sets the perpendicular width $\Delta x_\perp$ in the pump-probe response. Recall that the return amplitude is
\begin{equation}
    I(t_2)=\braket{2_p|e^{-iH t_2}|2_p}=\sum_\vec{r} \braket{2_p|e^{-iH (t_2-\tau)}|\vec {r}}\braket{\vec{r}|e^{-iH \tau}|2_p}
\end{equation}
where we insert an intermediate time $0<\tau<t_2$, at which the planon-pair state $\ket{\vec{r}}$ is represented by the difference coordinate ($\vec{r}=\vec{r}_1-\vec{r}_2$). For simplicity, we first consider the free propagator in two dimensions:
\begin{align}
  \braket{\vec{r'}|e^{-iHt}|\vec{r}}
    = \frac{m}{2\pi i  t} e^{i\frac{m|\vec r - \vec r'|^2}{2 t}} 
\end{align}
The initial state $\ket{2_p} = \frac{1}{2}(\ket{(1,0)}+\ket{(-1,0)}+\ket{(0,1)}+\ket{(0,-1)})$.
At late times and on length scales large compared with the lattice spacing ($r\gg 1$), the free propagator from $\ket{2}$ to $\ket{\vec r}$ can be approximated as
\begin{align}
    \braket{\vec{r}|e^{-iH t}|2_p} &=\frac{m}{2\pi i  t} e^{i\frac{m(|\vec r |^2+1)}{2 t}} \left[ \cos\frac{mr_x}{t}+\cos\frac{mr_y}{t}\right] \nonumber \\
   & \approx \frac{m}{\pi i  t} e^{i\frac{m|\vec r |^2}{2 t}}
\end{align}
where in the last step we used $mr/t\ll1$ at late times. It then follows that 
\begin{align}
      I(t_2) = -\frac{m^2}{\pi^2}\frac{1}{\tau(t_2-\tau)}\int d^2\vec r   e^{i \alpha(t_2, \tau) |\vec r|^2}
\end{align}
where 
\begin{equation}
    \alpha(t_2, \tau) = \frac{m}{2} \left(\frac{1}{\tau} + \frac{1}{t_2-\tau} \right)
    \label{eqn:r-integral}
\end{equation}
In the 2D spatial integral, contributions from separations satisfying $\alpha(t_2,\tau)|\vec r|^2>1 $ oscillate rapidly and mostly cancel. Thus, the typical distance is set by $|\vec{r}|_{\text{typ}}^2\sim 1/\alpha(t_2,\tau)$:
\begin{equation}
    \abs{\vec{r}}_{\text{typ}}^2 \sim \frac{2}{m} \frac{\tau(t_2-\tau)}{t_2}
    \label{eqn:alpha}
\end{equation}
Near the bottom band edge, the planon's band dispersion can be approximated as $\varepsilon_p\approx2h_zk^2\approx k^2/2m$ with the effective mass $m= 1/4h_z$. The maximum typical separation occurs at mid-time $\tau = t_2/2$, which yields
\begin{equation}
    \abs{\vec{r}}_\text{max} =\sqrt{2h_zt_2} 
\end{equation}
As a result, the planon perpendicular width scales as $ \Delta x_\perp  \sim O(\sqrt{h_zt_2}$) in the noninteracting case. 

We now comment on the modification due to the hard-core constraint. The propagator can be written in the energy basis as 
\begin{align}
    K (\vec r, t) &= \braket{\vec r| e^{-iHt}|2_p}= \int dE \; \rho(E)  \braket{\vec r|E}\braket{E|2_p} e^{-iEt}
\end{align}
In the presence of hardcore constraint, we have shown that $\braket{E|2_p} \propto \sqrt{k}/\log k $ near the band edge, and $\rho(E) \propto 1/k$ for the $l=0$ sector. In addition, 
\begin{align}
\langle \vec r|E\rangle
=2C_0(k)\Big[J_0(kr)-\tan\delta_0(k)\,Y_0(kr)\Big].
\end{align}
where $C_0(k) =\sqrt{k/2R}$ and $
\tan\delta_0(k)
=\mathcal O\!\left(\frac{1}{\ln k}\right)$ as $k \rightarrow 0$. In the energy integral at late times, change of variable implies the dominant momenta scale as $k\propto 1/\sqrt{t}$, while the typical separations satisfy $r\propto \sqrt{t}$ in the noninteracting case, so that $J_0(kr)$ and $Y_0(kr)$ remain $O(1)$ functions. We therefore make the approximation:
\begin{equation}
    \langle \vec r | E \rangle \approx 2C_0(k) J_0(kr)\left[ 1+i O\left(\frac{1}{\log k}\right)\right]
\end{equation}
where the imaginary correction $i O\left(\frac{1}{\log k}\right)$ arises because $J_0(x)$ and $Y_0(x)$ have an approximate $\pi/2$ phase difference if $x$ is not small.  Compared with the $l=0$ sector in the non-interacting case where $ \braket{\vec r|E}= 2C_0(k) J_0(kr)$ and $\braket{E|2_p} \approx 2C_0(k)\propto \sqrt{k}$, we can relate the free propagator $K_0(\vec r, t)$ to the propagator with hardcore constraint $K_{hc}(\vec r,t)$ as
\begin{align}
K_{\mathrm{hc}}(\vec r,t) & =\frac{A}{\log t}\,K_0( \vec r,t)\left[1+ i O\left(\frac{1}{\log t}\right)\right] \\
& \approx \frac{mA}{\pi i t\ln t} \exp{\left[i\frac{m|\vec r |^2}{2 t} + iO\left(\frac{1}{\log t}\right)\right ]}
\end{align}
where $A$ is a constant. The interaction therefore primarily modifies the propagator amplitude by a logarithmic factor, as expected. However, the planon’s typical width is determined by the oscillatory phase, which picks up only an overall phase shift that does not affect the relevant scale  $m|\vec r|^2/2t$. Following a similar derivation to Eqs. (\ref{eqn:r-integral}) and (\ref{eqn:alpha}), we therefore conclude that the planon typical width still scales as $\Delta x_\perp\propto \sqrt{h_zt_2}$ in the presence of the hardcore constraint. 

The derivation of the lineons' typical width proceeds in parallel with the planon case. In the noninteracting limit, the result is straightforward since the one-dimensional propagator retains the usual $e^{imx^2/2t}$ exponential factor, which immediately implies the typical width scaling $x\sim\sqrt{t}$. For interacting case, we first express the propagator as
\begin{align}
    K(x,t)
    &=
\braket{x|e^{-iHt}|2_\ell}
    =
    \int_0^\infty dk \,
    \braket{x|k}
    \braket{k|2_\ell}
    e^{-i\varepsilon_\ell(k)t},
\end{align}
where $\ket{2_\ell}$ is the nearby lineon-pair state, and $\varepsilon_\ell(k)\approx 2h_xk^2$ at low energies. The hardcore constraint gives $
\braket{k|2_\ell}\propto \sin k
$
and
$
\braket{x|k}\propto \sin(kx).
$
Up to an overall constant, the resulting propagator becomes
\begin{equation}
K_\text{hc}(x,t)
\propto
\sqrt{\frac{ 1}{h_zt}}
\,
e^{i (x^2+1)/(8h_zt)}
\sin\!\left(\frac{ x}{4h_z t}\right).
\end{equation}For simplicity, we set the intermediate time to $\tau=t_2/2$. The return probability can then be written as
\begin{align}
    I(t_2)
    &=
    \int dx \,
    \braket{2_\ell|e^{-iHt_2/2}|x}
    \braket{x|e^{-iHt_2/2}|2_\ell}
    \\
    &\propto
    \frac{1}{t_2}
    \int dx \,
    e^{i(x^2+1)/(4h_zt_2)}
    \sin^2\!\left(\frac{x}{4h_zt_2}\right).
\end{align}
At late times, the typical width is still determined by the oscillatory factor $e^{ih_zx^2/4t_2}$, leading to the scaling
$
x\sim O(\sqrt{h_zt_2}).
$
Using this scaling, the sine factor can be approximated as
$
\sin(x/4h_zt_2)\sim O(1/\sqrt{h_zt_2})
$ at late times. 
Applying the Fresnel integral then gives
\begin{align}
    I(t_2)
    &\propto
    \frac{1}{t_2^2}
    \int dx \,
    e^{ix^2/(4h_zt_2)}
    \propto
    \frac{1}{t_2^{3/2}}.
\end{align}
This result matches the lineon's linear-response signal $\chi_X^{(1)}(t_2)$ derived in Eq.~\ref{eqn:Lineon_linear_response}, as expected.

\section{Lineon excitation by the pump pulse \label{Appendix_F}}

We examine perturbing the system with a time-dependent magnetic field to excite a distribution of lineons pairs. Consider subjecting the system to a pulse of the form
\begin{equation}
    V(t) = B(t) \sum_{e} X_e
\end{equation}
where the field $B(t)$ assumes the form of a Gaussian wave packet of width $\tau$ and frequency $\omega$:
\begin{align}
    B(t) &= B_0  \exp\left(- \frac{t^2}{2\tau^2}\right) \cos\omega t \\
    &= \frac{1}{2} B_0 \exp\left(- \frac{t^2}{2\tau^2}\right) e^{i\omega t} + \text{c.c.}
    \label{eqn:pump-pulse}
\end{align}
We parametrize the frequency $\omega$ as $\omega = 2\Delta_\ell + \delta$, i.e., with detuning $\delta$ relative to the energy $2\Delta_\ell$ to create an isolated pair of lineons. The width $\tau$ is chosen such that the wave packet contains many oscillations, $\omega \tau \gg 1$. In the following, we study the lineon distribution as a function of the parameters entering Eq.~\eqref{eqn:pump-pulse}.

The state of the system at time $t$ is given within first-order time-dependent perturbation theory by
\begin{equation}
    \braket{m | \psi_I(t)} = -i \int_{-\infty}^{t} ds \, e^{i(E_m - E_n)s} B(s) \sum_e (X_e)_{mn} 
    \label{eqn:coefficients-interaction-picture}
\end{equation}
where $(O)_{mn} \equiv \braket{ m | O | n}$.
We then make use of the integral
\begin{align}
    \int_{-\infty}^{t} ds \, e^{i\omega s - s^2/2\tau^2} &= \int_{-\infty}^{t} ds \, e^{ - (s-i\omega \tau^2)^2/2\tau^2 - \omega^2\tau^2/2} \notag \\
    &\approx \sqrt{2\pi} \tau e^{- \omega^2\tau^2/2}
\end{align}
where the approximation holds for $t \gg \tau$, which allows us to extend the upper limit of integration to infinity. Performing the integral in Eq.~\eqref{eqn:coefficients-interaction-picture} and considering an eigenstate with two lineons along the $z$-oriented line at position $(x, y)$
\begin{multline}
    \braket{x, y; K_z, k_z | \psi_I(t)} \approx \\
    -i B_0 \tau \sqrt{\frac{\pi}{2}} \left[ e^{-(E_{mn} + \omega)^2\tau^2 / 2} + e^{-(E_{mn} - \omega)^2\tau^2 / 2} \right]  \\
    \times \sum_e \bar{A}(K_z, k_z) e^{-iK_z R_e} \sin k_z
\end{multline}
where $E_{mn} = E_m - E_n$.
The translation-invariant sum over edges implies that only lineon states with zero center of mass quasimomentum $(K_z=0)$ are excited. For such states,
%
%
%
%
\begin{multline}
    \braket{x, y; 0, k_z | \psi_I(t)} \approx \\
    -i  B_0 \tau \sqrt{2\pi} \left[ e^{-(E_{mn} + \omega)^2\tau^2 / 2} + e^{-(E_{mn} - \omega)^2\tau^2 / 2} \right]  \sin k_z
\end{multline}
Finally, we neglect the term that is not resonant. Since $E_m-E_n+\omega \approx 2\omega$,
and we have assumed that $\omega \tau \gg 1$, this term is suppressed as $e^{-2(\omega \tau)^2} \ll 1$. This leaves us with
\begin{equation}
    \braket{x, y; 0, k_z | \psi_I(t)} \approx -i  B_0 \tau \sqrt{2\pi}  e^{-[\varepsilon_\ell(k_z) - \delta]^2\tau^2 / 2}   \sin k_z
\end{equation}
where $\varepsilon_l$ is the energy relative to the bottom of the band. We assume both $\delta$ and $\tau^{-1}$ are small compared to the lineon bandwidth (i.e., $h_x \gg \delta, \tau^{-1}$), so it is appropriate to expand the lineon pair energy $\varepsilon_\ell(k)$ near the bottom of the band, giving $\varepsilon_\ell(k) = 2h_xk^2 + O(k^4)$. In the following, we discuss the two parameter regimes $\delta\tau \gg 1$ and $\delta\tau \ll 1$ separately.

Lineon excitations affect the braiding probability in two ways: through the lineon velocity and through the lineon excitation density. In the main text, we assume that the lineon pair moves with a characteristic speed $|v|$, but in general the excited lineons have a velocity distribution $p(v)$, and the braiding probability is modified to 
\begin{equation}
    P\big[\vec r_{1,2}(t); v]
  \;\propto\;
  \int\! dv\; p(v)\;
  A\big[\vec r_{1,2}(t);v]
  .
\end{equation}
The distribution $p(v)$ modifies the overall  pump-probe coefficient but does not affect the long-time scaling with $t_2$. For $\delta\tau\gg 1$, the excitations are concentrated on the resonant shell $\varepsilon_\ell(k_*)=\delta$ with
    $
      k_*=\sqrt{{\delta}/({2h_x})}$. The characteristic velocity is then 
      \begin{equation}
             v(k_*)=
             \left. \frac{\partial\varepsilon_\ell}{\partial k} \right\rvert_{k^*}
             = 2\sqrt{2h_x\,\delta}
             \, ,
      \end{equation}
which is set by the detuning $\sqrt \delta$. The width of the distribution $p(v)$  is controlled by the pulse duration and is of order $\tau^{-1}$, and in this regime it is typically smaller than $v(k_\star)$. In the opposite limit $\delta\tau\ll 1$,there is no sharply defined resonant shell, and all states with $\varepsilon_\ell(k)\,\tau\lesssim 1$ have appreciable excitation weight. This leads to a characteristic width of $p(v)$ that scales as $\tau^{-1/2}$. Explicitly, the characteristic momentum is set by $k_*\approx 1/\sqrt{2h_x\tau}$ and the characteristic velocity becomes 
\begin{equation}
    v(k_*)\approx2\sqrt{2h_x/\tau}
    .
\end{equation}

The braiding probability is proportional to the lineon density as long as the excitation density remains sufficiently low. Near the band edge, the lineon density of states exhibits a van Hove singularity (VHS), $g(\varepsilon_\ell)\propto1/\sqrt{\varepsilon}_\ell$, since only the mode with quasimomentum $K_z=0$ is excited. However, the matrix element vanishes at the band edges as $|c_k|^2 \propto (\sin k)^2 \sim k^2 \sim \varepsilon_\ell$. The lineon density can be evaluated as $\rho=\int \frac{dk}{2\pi} |c_k|^2 = \int d\varepsilon g(\varepsilon)|c_k|^2$  in two regimes. For $\delta\tau\ll 1$,
\begin{equation}
    \rho = \frac{B^2_0}{4 h_x} \Gamma(3/4) \tau^{1/2} \left[1+2\frac{\Gamma(5/4)}{\Gamma(3/4)} (\delta\tau) +O(\delta^2\tau^2) \right]
\end{equation}
For $\delta\tau\gg 1$,
\begin{equation}
    \rho = \frac{\sqrt{\pi}B^2_0}{2 h_x} \tau \delta^{1/2} \left[1 + O((\tau \delta)^{-2})   \right]
    .
\end{equation}
Therefore, the expected VHS at the band edge is regularized in the excitation density due to the hardcore constraint.
Combining the results for the lineon density with the characteristic velocity and keeping only the leading order, we obtain (at a fixed time $t_2$):

\begin{equation}
\chi_\mathit{ZZX}(\tau, \delta) \propto 
\begin{cases}
    \chi_Z^{(1)}\times \tau^0\delta^0, &\delta\tau\ll 1\\[2pt]
   \chi_Z^{(1)}\times \tau\delta,  &\delta\tau\gg 1\\[2pt] 
\end{cases}
\end{equation}
where $\chi_Z^{(1)}$ is the linear-response susceptibility of the planons created by the probe pulse, while $\tau$ and $\delta$ are pump parameters that control the excitation of lineons. Although the VHS in the lineon's DOS does not show up, $\chi_Z^{(1)}$ might have a VHS at half filling due to the effective 2D tight-binding Hamiltonian of planons. 

\section{Time scales for $t_1$}
Take the ZZX protocol as an example, in order for the planon pair to braid only one of the two lineons, the separation between the two pumped lineons should be larger than the typical spatial width of the planon pair during the probe evolution. For planon's bound state, we only require $vt_1 \sim O(1)$.
The characteristic velocity of the lineons is
$v=2\sqrt{2h_x\delta}$ for $\delta\tau\gg 1$, and
$v=2\sqrt{2h_x/\tau}$ for $\delta\tau\ll 1$. For the extended states, the typical width reaches its maximum value $\sim O(\sqrt{h_z t_2})$ near the middle of the probe evolution. At the same time, the lineon pair has propagated for a time interval $t_1+t_2/2$. Therefore, we require
\begin{equation}
    2v\left(t_1+\frac{t_2}{2}\right) >  O(\sqrt{h_z t_2}).
\end{equation}
This relation holds for any sufficiently late time $t_2$. Nevertheless, we can impose a stricter condition on $t_1$, namely
\begin{equation}
    2vt_1 > O(\sqrt{h_z t_2}).
\end{equation}
This gives
\begin{equation}
    t_1 > O\left(\sqrt{\frac{h_z \tau t_2}{h_x}} \right),
    \qquad
    \delta\tau\ll 1,
\end{equation}
and
\begin{equation}
    t_1 > O\left(\sqrt{\frac{h_z  t_2}{\delta h_x}} \right),
    \qquad
    \delta\tau\gg 1.
\end{equation} 
In the XXZ protocol, we only need to deal with the extended states contribution, and the criterion for $t_1$ can be estimated by simply reversing $h_x$ and $h_z$.

\bibliography{biblio}

\end{document}